\documentclass{nature}
\usepackage{textcomp}
\usepackage[a4paper,top=2cm,bottom=2cm,left=2.cm,right=2.cm]{geometry}  
\usepackage{epsfig}
\usepackage{emptypage}
\usepackage{babel}
\usepackage[bf, footnotesize]{caption}
\usepackage{color}
\usepackage{amsmath}
\usepackage{amssymb}
\usepackage[T1]{fontenc}
\usepackage[dvipsnames]{xcolor}
\definecolor{gray}{gray}{0.25}
\usepackage{lastpage}
\usepackage{fancyhdr}
\usepackage{hyperref}
\usepackage{lineno}
\usepackage{caption}
\usepackage{subcaption}

\title{A titanic interstellar medium ejection from a massive starburst galaxy at z=1.4}

\begin{document}

\pagenumbering{gobble} 

\author{Annagrazia Puglisi*$^{1, 2}$, Emanuele Daddi$^2$,  Marcella Brusa$^{3, 4}$, Frederic Bournaud $^2$, 
Jeremy Fensch$^5$, Daizhong Liu$^6$, Ivan Delvecchio$^2$, Antonello Calabr{\`o}$^{7}$, Chiara Circosta$^8$, 
Francesco Valentino$^{9}$, Michele Perna$^{10, 11}$, 
Shuowen Jin$^{12,13}$,  Andrea Enia$^{14}$, Chiara Mancini$^{14}$, Giulia Rodighiero$^{14, 15}$}

\maketitle

\begin{affiliations}
\item Center for Extragalactic Astronomy, Durham University, South Road, Durham DH1 3LE, United Kingdom %1
\item CEA, IRFU, DAp, AIM, Universit{\'e} Paris-Saclay, Universit{\'e} Paris Diderot, Sorbonne Paris Cit{\'e}, CNRS, F-91191 Gif-sur-Yvette, France %2
\item Dipartimento di Fisica e Astronomia, Universit{\`a} di Bologna, via Gobetti 93/2, 40129 Bologna, Italy %3
\item INAF-Osservatorio Astronomico di Bologna, via Gobetti 93/3, 40129 Bologna, Italy %4
\item Univ. Lyon, ENS de Lyon, Univ. Lyon 1, CNRS, Centre de Recherche Astrophysique de Lyon, UMR5574, F-69007 Lyon, France %5
\item Max Planck Institute for Astronomy, Konigstuhl 17, D-69117 Heidelberg, Germany %6
\item INAF-Osservatorio Astronomico di Roma, Via Frascati 33, I-00040 Monte Porzio Catone Roma, Italy %7
\item Department of Physics \& Astronomy, University College London, Gower Street, London WC1E 6BT, United Kingdom %8
\item Cosmic Dawn Center at the Niels Bohr Institute, University of Copenhagen and DTU-Space, Technical University of Denmark, Denmark %9
\item Centro de Astrobiolog\'ia (CAB, CSIC--INTA), Departamento de Astrof\'\i sica, Cra. de Ajalvir Km.~4, 28850 -- Torrej\'on de Ardoz, Madrid, Spain %10
\item INAF-Osservatorio Astrofisico di Arcetri, Largo Enrico Fermi 5, I-50125 Firenze, Italy %11
\item Instituto de Astrof{\'i}sica de Canarias (IAC), E-38205 La Laguna, Tenerife, Spain %12 
\item Universidad de La Laguna, Dpto. Astrof{\'i}sica, E-38206 La Laguna, Tenerife, Spain %13
\item Dipartimento di Fisica e Astronomia, Universit{\`a} di Padova, vicolo dellOsservatorio 2, I-35122 Padova, Italy %14 
\item INAF-Osservatorio Astronomico di Padova, Vicolo dell'Osservatorio, 5, I-35122 Padova, Italy %15
\end{affiliations}

\begin{abstract}
Feedback-driven winds from star formation or active galactic nuclei might be a relevant channel for the abrupt quenching star formation in massive galaxies. However, both observations and simulations support the idea that these processes are non-conflictingly co-evolving and self-regulating. Furthermore, evidence of disruptive events that are capable of fast quenching is rare, and constraints on their statistical prevalence are lacking. Here we present a massive starburst galaxy at $z=1.4$ which is ejecting $46 \pm 13$\% of its molecular gas mass  at a startling rate of $\gtrsim 10,000$ M$_{\odot}{\rm yr}^{-1}$. A broad component that is red-shifted from the galaxy emission is detected in four (low- and high-J) CO and [CI] transitions and in the ionized phase, which ensures a robust estimate of the expelled gas mass. The implied statistics suggest that similar events are potentially a major star-formation quenching channel. However, our observations provide compelling evidence that this is not a feedback-driven wind, but rather material from a merger that has been probably tidally ejected. This finding challenges some literature studies in which the role of feedback-driven winds might be overstated.
\label{Abstract}
\end{abstract}

The physical processes responsible for the sudden termination of star formation in massive galaxies have not been clarified yet\cite{ManBelli19}. 
Feedback-driven winds launched by star formation or powerful active galactic nuclei (AGN) have been often indicated as a viable channel to rapidly quench star formation in massive galaxies\cite{Harrison17, Cattaneo09}. 
From the point of view of theory, different flavours of galaxy evolution models\cite{Schaye15, Weinberger17} require  energy injection from AGN into the interstellar medium of galaxies 
to explain many observed properties of massive galaxies and the phenomenology of quenching. 
This includes the formation of massive red and dead elliptical galaxies, no longer forming stars at a significant rate. 

In line with the postulates from galaxy evolutionary models, observations of distant galaxies are often interpreted suggesting that feedback-driven winds are widespread among most massive galaxies\cite{ForsterSchreiber19}, exactly those that are expected to quench rapidly. 
Thus, there is a broad consensus on the idea that feedback-driven winds are essential for galaxy evolution and quenching. 

On the other hand, observations  demonstrate the existence of tight correlations between the properties of accreting black holes and their host galaxies\cite{Mullaney12, KormendyHo13}. Also, the history  of black hole growth and star-formation look remarkably similar across cosmic time\cite{MadauDickinson14}.
These empirical pieces of evidence all suggest that  accretion activity onto  central black-holes co-exists with star formation and that these two phenomena are  self-regulating over  billion years timescales corresponding to the evolution of their host\cite{Schreiber15}, rather than competing or being mutually exclusive.
This idea is also corroborated by numerical simulations suggesting that feedback-driven outflows have no immediate impact on star formation in massive galaxies\cite{GaborBournaud14}.

Studying feedback-driven outflows from a theoretical point of view is challenging because these processes involve  connecting phenomena occurring at very different temporal and spatial scales.
As such, direct observations of disruptive events unambiguously demonstrating the impact of feedback-driven winds on the host galaxy star formation termination are essential.
So far, such observations of extreme, smoking-gun events are rare, possibly limited to a single claim arising from an object with peculiar properties\cite{Geach14}.
It is thus difficult to interpret these observations in the context of the massive galaxies population and to infer conclusions on the prevalence of outflow-driven quenching and thus in the direct formation of giant, passively evolving early-type galaxies.

\section{A disruptive event in a massive galaxy at z=1.4}

ID2299 is a starburst galaxy at z = 1.395 that has a stellar mass M$_{\star} = 9.4 \pm  1.3 \times 10^{10}$ M$_{\odot}$ and is experiencing intense star-forming activity with a star formation rate SFR$ = 550 \pm 10$ M$_{\odot} {\rm yr}^{-1}$.
This is $\sim 5 \times$ higher than the average SFR of star-forming galaxies with comparable stellar mass at the same cosmic epoch\cite{Schreiber15}.
The starburst in this source is occurring in a compact region of effective radius R$_{\rm eff, ALMA} = 1.54 \pm 0.04 $ kpc (see Methods).
This galaxy hosts a black hole showing signs of ongoing accreting activity. The object is detected in the X-rays and has a weak rest-frame 2-10 keV absorption-corrected luminosity\cite{Civano16} L$_{\rm X-ray} \sim 1.6 \times 10^{43} \ {\rm erg s^{-1}}$ partially arising from the intense star-forming activity of the galaxy ($\sim 10 - 15 \%$ according to typical X-ray to SFR conversions\cite{Lehmer16}).

This source was selected from a large Atacama Large Millimeter/sub-millimeter Array (ALMA) survey observing the carbon monoxide J = $2 \rightarrow 1$, $5 \rightarrow 4$ and $7 \rightarrow 6$ rotational transitions and the neutral atomic carbon transition [CI](2-1) in 123 far-IR selected galaxies at $1.1 \leqslant z \leqslant 1.7$ (see Methods). 
The CO(2-1) and [CI](2-1) transitions trace the bulk of the cold molecular gas reservoir while the CO(5-4) and CO(7-6) transitions sample the dense molecular gas. 
The emission from the galaxy is detected at 15 and 28 $\sigma$ in CO(2-1) and CO(5-4) respectively and at more than $50 \ \sigma$ in CO(7-6) and [CI](2-1). These lines have a full width half maximum v$_{\rm FWHM, n} = 120 \pm 13 \ {\rm km s^{-1}}$ at $z_{\rm sub-mm} = 1.395 \pm 0.017$, in excellent agreement with the redshift obtained from rest-frame optical emission lines (Figure \ref{Fig1_spectra1d}, Table \ref{Table}).
The CO(2-1) and CO(5-4) spectra reveal the presence of a broad component at $6.2 \ \sigma$ and $3.5 \ \sigma$ significance, respectively (see Methods).
This feature is detected at 2.7 $\sigma$ in [CI](2-1), implying  CO(2-1)/[CI](2-1) ratios consistent with what observed in distant galaxies \cite{Valentino19}, while CO(7-6) provides 
 an upper limit.
The broad component has v$_{\rm FWHM, b} = 535 \pm 135 \ {\rm km s^{-1}}$ and its centroid is $179 \ \pm 78 \ {\rm km s^{-1}}$ red-ward of the systemic velocity of the galaxy identified by the centroid of the narrow emission. 
The broad emission shows no significant spatial offset with respect to the galaxy (Figure \ref{Fig2_maps}, see also Methods).
A broad component is also detected at more than $15 \sigma$ significance around the singly ionized oxygen forbidden emission ${\rm [OII]_{3726, 3729}}$  in the rest-frame optical spectrum from the DEIMOS 10K Spectroscopic survey\cite{Hasinger18} (Figure \ref{Fig1_spectra1d}).
The ${\rm [OII]_{3726, 3729}}$ broad components are detected at {$92 \pm 47 \ {\rm kms^{-1}} $} separation from the narrow lines and have v$_{\rm FWHM, b} = 537 \pm 106 \ {\rm km s^{-1}}$, consistently to what measured on the ALMA spectra. 
The remarkable consistency of the kinematics in the ionized and molecular phases implies that these are tracing different phases of the same phenomenon.

We measure the molecular gas mass of the system from the CO(2-1) luminosity (Methods). We assume a starburst-like CO excitation (L$^{'}_{\rm CO(2-1)}$/L$^{'}_{\rm CO(1-0)} = 0.85$) and a CO-to-$H_2$ conversion factor $\alpha = 0.8 \ {\rm M}_{\odot}({\rm K \ km \ s^{-1} pc^2})^{-1}$ which is appropriate for the interstellar medium (ISM) conditions of the galaxy. 
We derive M$_{\rm mol, n} = 2.3 \pm 0.2 \times 10^{10}$ M$_{\odot}$ and M$_{\rm mol, b} = 2.0 \pm 0.5 \times 10^{10}$ M$_{\odot}$ for the narrow and broad component respectively. 
{Therefore, $46 \pm 13 \%$ of the total molecular gas mass is decoupled from the galaxy.}
This is a conservative estimate given the lower CO excitation in the broad component (Methods).

\section{Impact on the host galaxy}

Assuming that the gas in the broad component is associated to a feedback-driven wind (e.g., from AGN or star-formation), would translate its gas mass into an exceptional mass outflow rate $\dot{\rm M}_{\rm mol, OF} = 25000 \ \pm \ 8000 $ M$_{\odot}{\rm yr}^{-1}$. 
{Consistently with the literature definitions \cite{Fiore17} (see Methods), } this computation assumes that the outflow has a velocity v$_{\rm max, OF} = 634 \pm 139 \ {\rm km s^{-1}}$ and extends over R$_{\rm OF} = 1.54 \pm 0.04$ kpc corresponding to the radius over which the ALMA spectra are extracted.
The velocity and spatial extent of the outflow imply a  propagation timescale t$_{\rm OF} = {\rm R}_{\rm OF} / {\rm v}_{\rm max} = 3 \pm 1 $ Myr
{ (t$_{\rm OF} \sim 10$ Myr and $\dot{\rm M}_{\rm mol, OF} = 7100 \ \pm \ 3600 $ M$_{\odot}{\rm yr}^{-1}$, if one were instead to use the most conservative average velocity shift v$_{\rm mol, b} = 179 \ {\rm km s^{-1}}$ )}.
This suggests that such enormous quantity of gas (roughly half of the original galaxy ISM, mostly composed by molecular gas at these cosmic epochs \cite{Daddi10}) is being expelled through a quasi-instantaneous event, not accumulating over the Gyr-long timescales of galaxy activity. 

This disruptive event, including a starbursting core and with half or more of the molecular gas mass being expelled, will eventually lead the galaxy to quench. 
{ The range of outflow velocities (likely a lower limit to the intrinsic distribution of ejection velocities because of the unknown inclination) reaches higher than the galaxy escape velocity (v$_{\rm esc} = (2G{\rm M}_{\star}/{\rm R}_{\rm eff, K_{s}})^{1/2} \sim 500$ km/s at the K$_{\rm s}$-band galaxy effective radius, following Reference \cite{Emonts15}). This shows that at least part of the broad component material is able to escape the galaxy potential.} If this gas  proceeds at a constant speed, it will reach the virial radius of the system r$_{\rm vir} \sim 200$ kpc\cite{Springel08} in a few hundred~Myr. Part of this material could fall back at later times \cite{Falgarone17}, possibly rejuvenating the remnant\cite{Mancini19,Carnall19}.
The gas not escaping the galaxy potential is nonetheless ejected from the starburst region.
Thus, most of the material in the broad component will not be available for  star formation for several hundred Myr.
At the observed SFR, the gas remaining in the host would be consumed by star formation in $\tau_{\rm depl, SB} =$ M$_{\rm mol, n} / $ SFR $\sim 40$ Myr. 
Therefore, the residual material will be rapidly consumed long before the expelled gas could be re-accreted.
Furthermore, additional processes such as virial shocks or radio AGN activity would heat up the gas inhibiting its re-accretion at later times.
We are thus witnessing a system in which a typical massive galaxy is about to be quenched (and to remain such for a substantial amount of time), as a result of a violent episode having deprived it of a large fraction of its ISM content. 

\section{Frequency of disruptive events and connection with the quenched galaxy population}

{The host is experiencing a major merger (see Methods) that will rapidly evolve into a  quenched spheroid as suggested by simulations \cite{DiMatteo05} and by the starburst size which is consistent with typical early-type galaxies sizes at $z \sim 1.4$\cite{Puglisi19}}.
This single event was extracted from a survey of 123 objects and  thus implies $\sim (1 - 2) \times 10^4 {\rm /Mpc^3/Gyr}$ disruptive events per cosmic volume and unit time (Figure \ref{Fig5_number_density}, see Methods). The error associated to this measurement is $\sim 0.6 \ {\rm dex}$ reflecting the large uncertainty associated with a single event observation. 
However, the disruptive events statistics match within the error bars the expected density of newly quenched galaxies in the redshift range probed by our survey, based on the differential growth of quiescent galaxies as a function of redshift. 
This implies that this fast quenching channel is potentially capable to account for an important fraction of all newly quenched galaxies.

\section{Triggering mechanism of the disruptive event}

Broad components in galaxy spectra are typically interpreted as winds ejected by AGN/star-forming feedback activity. These winds are often blue-shifted albeit they can also appear red-shifted possibly as a result of the wind geometry and orientation with respect to the observer \cite{Perna15,Feruglio15,HerreraCamus19}.
Their energy depends on the power of the engine and the mechanism regulating the amount of energy transferred to the surrounding ISM.
The ID2299 mass outflow rate is inexplicable given the bolometric luminosity of the AGN embedded in the system and the strength of the starburst (Fig.~\ref{Fig3_scaling_rels}). This putative wind lies $\sim$ {2 (1.5)} orders of magnitudes above literature scaling relations\cite{Fiore17, Fluetsch19} implying a wind that would carry $\sim$ {2 (1.5)} orders of magnitudes more mass and energy than expected from the AGN (starburst) activity.
{Feedback-driven winds stronger than expected from the ongoing AGN/starburst activity have been interpreted as ``fossil outflows'' in which the powering engine
has faded by the time the outflow is observed \cite{Fluetsch19}.
However, the SFR $\geqslant 14000$ M$_{\odot}$yr$^{-1}$ implied by the scaling relation in the right panel of Figure \ref{Fig3_scaling_rels} cannot be ascribed to a single galaxy, as this is larger than the SFR of the brightest proto-clusters in the Universe\cite{Negrello20}.
Instead, the AGN bolometric luminosity can vary substantially over a $\sim 10^5$ yr time-scale\cite{Schawinski15}. 
However, in this case the 'just disappeared' AGN must have been as bright as L$_{\rm bol} \sim 10^{47.8} \ {\rm erg s^{-1}}$. The lack of bright radio emission already suggest that there is a low  probability that ID2299 was recently hosting such a powerful quasar (see Methods), whose extreme bolometric luminosity would approach that of the most luminous QSO known in the whole sky\cite{Wolf18}.}
Here we use state-of-the-art simulations of AGN-driven winds inside gas-rich massive galaxies\cite{GaborBournaud14} to further demonstrate that it is impossible to reproduce the event detected regardless of the underlying current or past AGN luminosities (Figure \ref{Fig4_OF_simulations}{, see also Methods}). 
Hence, the properties of this disruptive event cannot fit in the standard feedback-driven ejection interpretation. 

To better understand the triggering mechanism of this event, we can consider the clear merging nature of ID2299 (Methods). 
The velocity dispersion of the broad component is too large to be due to a merging galaxy companion because its implied dynamical mass would largely remain  unaccounted for (see Methods). This has therefore to be largely unbound gas. Various physical effects co-exist to drive complex velocity fields for the gas affected by the overall dynamics in  a merging starburst, including enhanced gas turbulence, torques-induced inflows and outflows \cite{Bournaud11}. This makes it challenging to unambiguously identify the mechanism producing the variety of gas kinematic components in merging starbursts integrated spectra (see Methods).
However, it is hard to reconcile the broad component of ID2299 with a kinematically perturbed disk since its excitation conditions are dramatically different than in the narrow component { (Figure \ref{Fig_LVGmodels}, see also Methods)} and this is at odds with observations of distant starbursting disks showing an overall high excitation\cite{Swinbank10, Danielson11}. 
A tidal ejection seems instead the most plausible scenario for explaining the bulk motion of roughly half of the total ISM in the system.

In fact, ID2299 is caught in a stage where merger-induced torques have caused the compaction\cite{BarnesHernquist} of large part of the molecular gas into a region that is $\sim 2 \times$ smaller than the stars that is currently experiencing a strong starburst (see Table \ref{Table} and Methods). 
Because of angular momentum conservation, this requires that, at the same time, a substantial amount of gas is also being expelled\cite{HibbardVanGorkom}. 
Indeed, mergers and interactions in the local Universe can eject large quantities of neutral gas through tidal tails with velocity dispersion of a few hundred km s$^{-1}$~\cite{Bournaud04}.
The tidally ejected material would be composed by denser molecular gas at high redshift since galaxies in the distant Universe are richer in molecular gas than their local counterparts\cite{Daddi10}.
This is directly observed in a recent ALMA study of a z=1.5 galaxy where tidally ejected material spreads to large distances accumulating already 20\% of the gas mass after the first pericentric passage \cite{Silverman18}.
Also hydrodynamic simulations of gas-rich mergers at $z \sim 2$ show that a non-negligible fraction of the molecular gas content of the system can be ejected through tidal tails in few mega-years with kinematic features consistent with our observations\cite{Fensch17}.
The tidally ejected gas in local mergers  shows signs of in-situ star formation that can contribute up to $\sim 15 - 20 \%$ to the total SFR of the system\cite{Bournaud10}. 
Here we can compute the SFR in the ejected material through the CO(5-4) emission tracing the dense phase of the molecular gas\cite{Daddi15}. From the luminosity of the broad CO(5-4) component, we estimate that the ejected material is forming stars at a rate of 
SFR$_{\rm broad} = 180 \pm 60$ M$_{\odot} {\rm yr}^{-1}$. This corresponds to $30 \pm 11 \%$ of the total SFR derived from the far-IR bolometric luminosity of the dust.
In addition, the molecular gas emission in the ejected component is significantly less excited than that of its host { (Figure \ref{Fig_LVGmodels})}. This is at odds with results from feedback-driven outflows physics while it is consistent with the tidal ejection hypothesis (see Methods).
Furthermore, the density of starburst galaxies capable of ejecting large tails of gas is consistent with the frequency of disruptive events (Figure \ref{Fig5_number_density}, see also Methods).
We thus conclude that the ID2299 observations can be explained by an exceptional tidal ejection episode triggered by the merger. {This does not affect  estimates of the frequency of these events (Fig.~\ref{Fig5_number_density}), given the conservative approach adopted (Methods), while allowing for less extreme inferences of outflowing mass rates of $\dot{\rm M}_{\rm mol, OF} = 7100 \ \pm \ 3600 $ M$_{\odot}{\rm yr}^{-1}$ if  using the average broad component velocity shift (see Methods).}

\section{Outflows triggering mechanisms mis-identification ?}

Tidally ejected material expelled during mergers can present observational features quite similar to feedback-driven winds. Hence, the two phenomena can be confused. 
Our observations trace an extreme episode  unlikely to be consistent with the classical interpretation of a feedback-driven wind. It is thus plausible that further, less extreme tidal ejection events might have gone misinterpreted as feedback-driven outflows in the literature. 
{For example, if we had only measured ID2299's ionized outflow energetics from ${\rm [OII]_{3726, 3729}}$ we might have interpreted it as a feedback-driven event, as often done in the distant Universe (see Methods).
Recently, star formation has been observed within a blue-shifted feature interpreted as a AGN/starburst-driven wind in a local ultra luminous infrared galaxy \cite{Maiolino17}}. This object is an advanced-stage merger showing extended tidal tails in the HST imaging\cite{Leslie14}. The outflowing material has velocity and velocity dispersion of a few hundred  km~s$^{-1}$, resembling the features of tidally ejected material which can also produce high velocity shocked regions\cite{MonrealIbero10}. The presence of star formation within this blue-shifted feature would be less surprising in a tidal ejection interpretation. 
Similarly, the overall properties (velocity field, spatial extent, morphology, etc.) of massive outflows discovered in a number of local and distant sources\cite{Cicone18,Sakamoto14,Geach14} resemble those observed in tidal tails\cite{Bournaud04}, where the large velocity dispersion might be explained by the high gas fraction of the host increasing the gas turbulence and clumpiness\cite{Bournaud07} as well as by the large spread of outflowing directions.
Extended tidal tails have been indeed identified in some such objects \cite{English03, Feruglio13} (see also Methods).
Finally, we note that the  incidence of putative feedback-driven outflows in distant galaxies increases as a function of the specific SFR \cite{ForsterSchreiber19, Swinbank19, Ginolfi20}, similarly to the frequency of mergers and interactions\cite{Cibinel19}.
We are not questioning that feedback-driven outflows do play a role. They are required, e.g., to explain the highest velocity components in galaxy and AGN spectra. However, their prevalence might have been overestimated and at least part of the literature on outflows and their overall impact on galaxy evolution might be worth reconsidering. 
This is perhaps best illustrated by the strong tension between the trends reported in Fig.~\ref{Fig3_scaling_rels} and Fig.~\ref{Fig4_OF_simulations}, corresponding to the current mainstream interpretation of observations versus state-of-the-art modelling of the same phenomenon (see also Methods).
Tidal ejections should be systematically considered when interpreting multiple kinematic components in galaxy spectra, particularly for merging or starbursting galaxies.

\newpage

\begin{addendum}
\item[Acknowledgements:] 
A.P. and E.D. thank Alvio Renzini for commenting the manuscript and for useful discussion.
AP acknowledges funding from Region \^{I}le-de-France and Incoming CEA fellowship from the CEA-Enhanced Eurotalents program, co-funded by FP7 Marie-Skłodowska-Curie COFUND program (Grant Agreement 600382).
A.P. also gratefully acknowledges financial support from STFC through grants ST/T000244/1 and ST/P000541/1.
M.P. acknowledges support from the Comunidad de Madrid through the Atracción de Talento Investigador Grant 2018-T1/TIC-11035
S.J. acknowledges financial support from the Spanish Ministry of Science, Innovation and Universities (MICIU) under grant AYA2017-84061-P, co-financed by FEDER (European Regional Development Funds). 

\item[Author contributions:]
A.P. and E.D. reduced the data, interpreted the results and wrote the paper. All other authors contributed to the observing proposals, to the scientific discussion and elaboration of the results, and provided comments to the manuscript. 

 \item[Correspondence] Correspondence and requests for materials
should be addressed to Annagrazia Puglisi ~(email: annagrazia.puglisi@durham.ac.uk).

\end{addendum}

%\newpage

\begin{figure}
\centering
\includegraphics[scale=0.8]{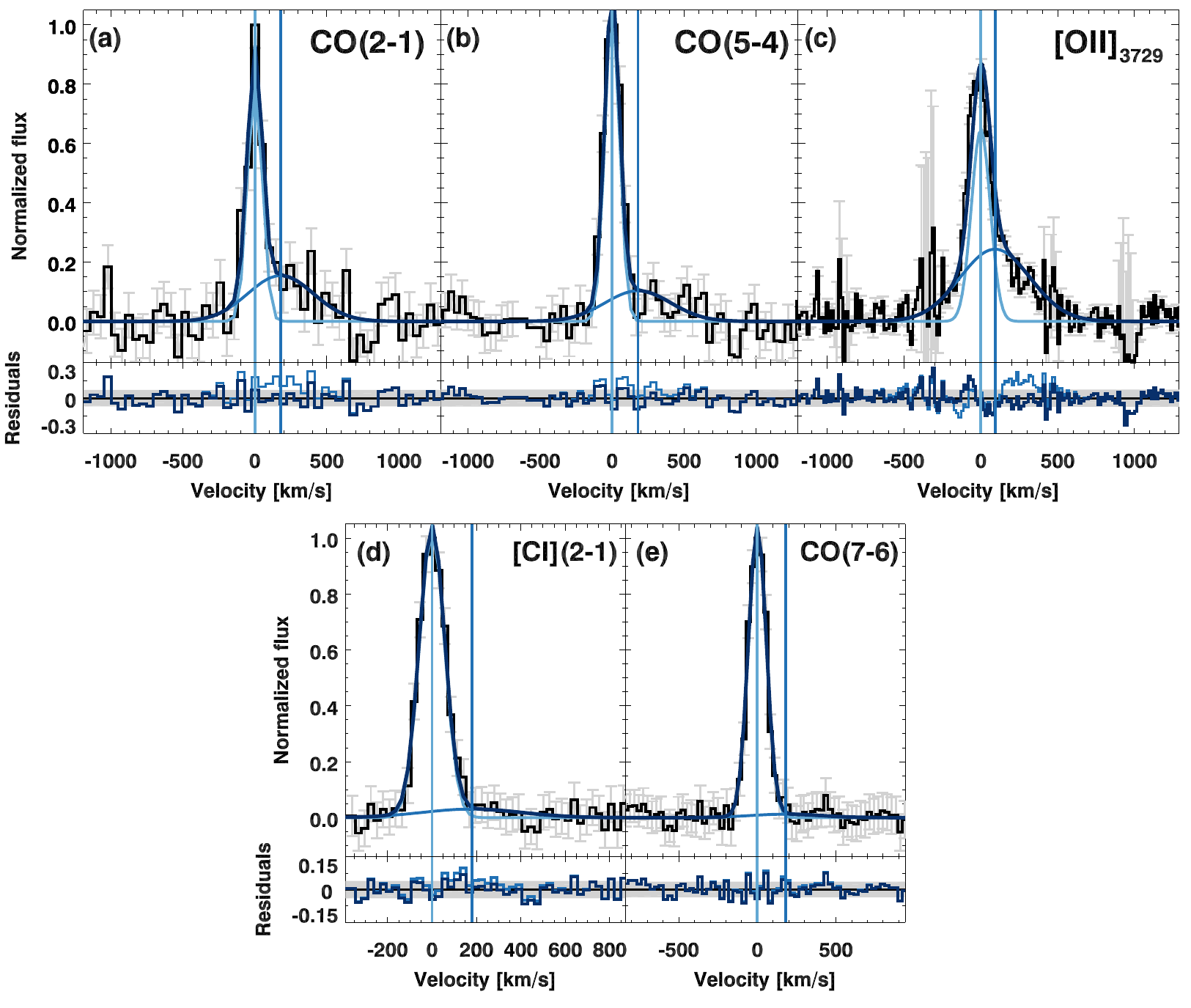}
\caption{ {\bf Multi-wavelength spectra of ID2299.}
Panel a-e: CO(2-1), CO(5-4), ${\rm [OII]_{3729}}$, [CI](2-1) and CO(7-6) spectrum of the source.
To aid the visual comparison with the CO/[CI] spectra, in panel c the $\lambda_{\rm rest} = 3726$ {\AA} narrow+broad Gaussian model has been subtracted from the observed spectrum and only the $\lambda_{\rm rest} = 3729$ {\AA} narrow+broad complex of the ${\rm [OII]_{3726, 3729}}$ doublet is shown.
The spectra are continuum-subtracted and are centred to the systemic velocity of the galaxy. 
The spectra are also normalized to the peak of the narrow emission.
In all top panels the black curve and the light grey errorbars represent the observed flux and 1$\sigma$ uncertainty, respectively.
Here, light blue lines show the Gaussian fits to the narrow and broad component. The darkest blue curve represents the total fit. 
Solid vertical lines mark the centroids positions. 
Below each spectrum, light and dark blue curves show relative residuals from a single and a two Gaussian fit, respectively. 
Here, light gray horizontal areas highlight the RMS level of the single Gaussian fit residuals.}
\label{Fig1_spectra1d}
\end{figure}

\begin{figure}
\centering
\includegraphics[scale=1.2]{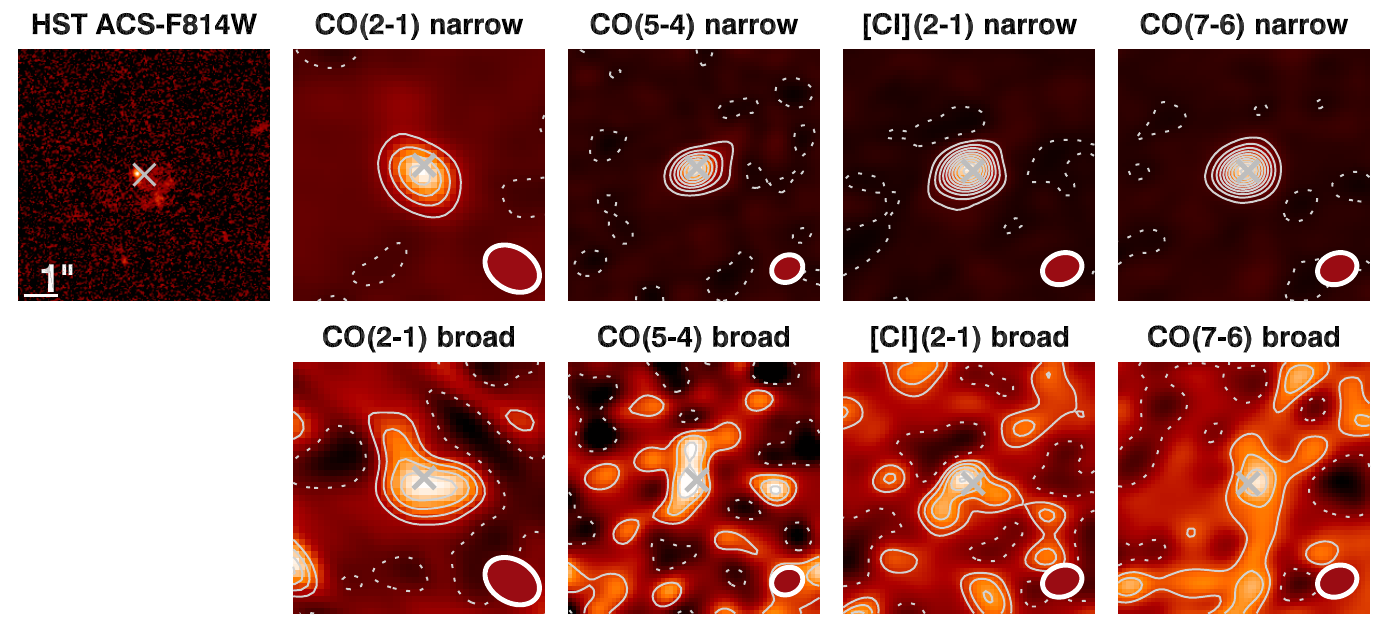}
\caption{   {\bf HST imaging and narrow and broad components ALMA maps of ID2299. }
The top-left panel shows the HST-F814W imaging of the source, sampling the UV rest-frame emission from young stars.
The top (bottom) rows show the CO(2-1), CO(5-4), [CI](2-1) and CO(7-6) ALMA maps of the narrow (broad) emission.
In all panels, the grey cross indicates the galaxy position in the K$_{\rm s}$-band, tracing the stellar continuum. All cut-outs have a size of $7 ''$. 
The beams are shown in the bottom right corner of each ALMA map.
The narrow component maps have been created by collapsing the ALMA cubes within a velocity range corresponding to 3$\times$ the line FWHM. Contours starts at $\pm 1 \sigma$ and go in steps of $3 \sigma$.
The broad component maps have been obtained by collapsing the continuum and narrow component-subtracted cube in a velocity range corresponding to the line FWHM with the exception of the CO(2-1) broad component map that has been collapsed in the 3$\times$ FWHM velocity range. Contours starts at $\pm 1 \sigma$ and go in steps of $1 \sigma$. 
Negative contours are highlighted with dashed curves in each ALMA map.
From the top left, the RMS of each narrow component ALMA map is $0.36$ mJy beam$^{-1}$; $0.67$ mJy beam$^{-1}$; $0.18$ mJy beam$^{-1}$; $0.18$ mJy beam$^{-1}$. 
The RMS of each broad component map is $0.21$ mJy beam$^{-1}$; $0.41$ mJy beam$^{-1}$; $0.11$ mJy beam$^{-1}$; $0.12$ mJy beam$^{-1}$.}
\label{Fig2_maps}
\end{figure}

\begin{figure}
\centering
\includegraphics[scale=0.8]{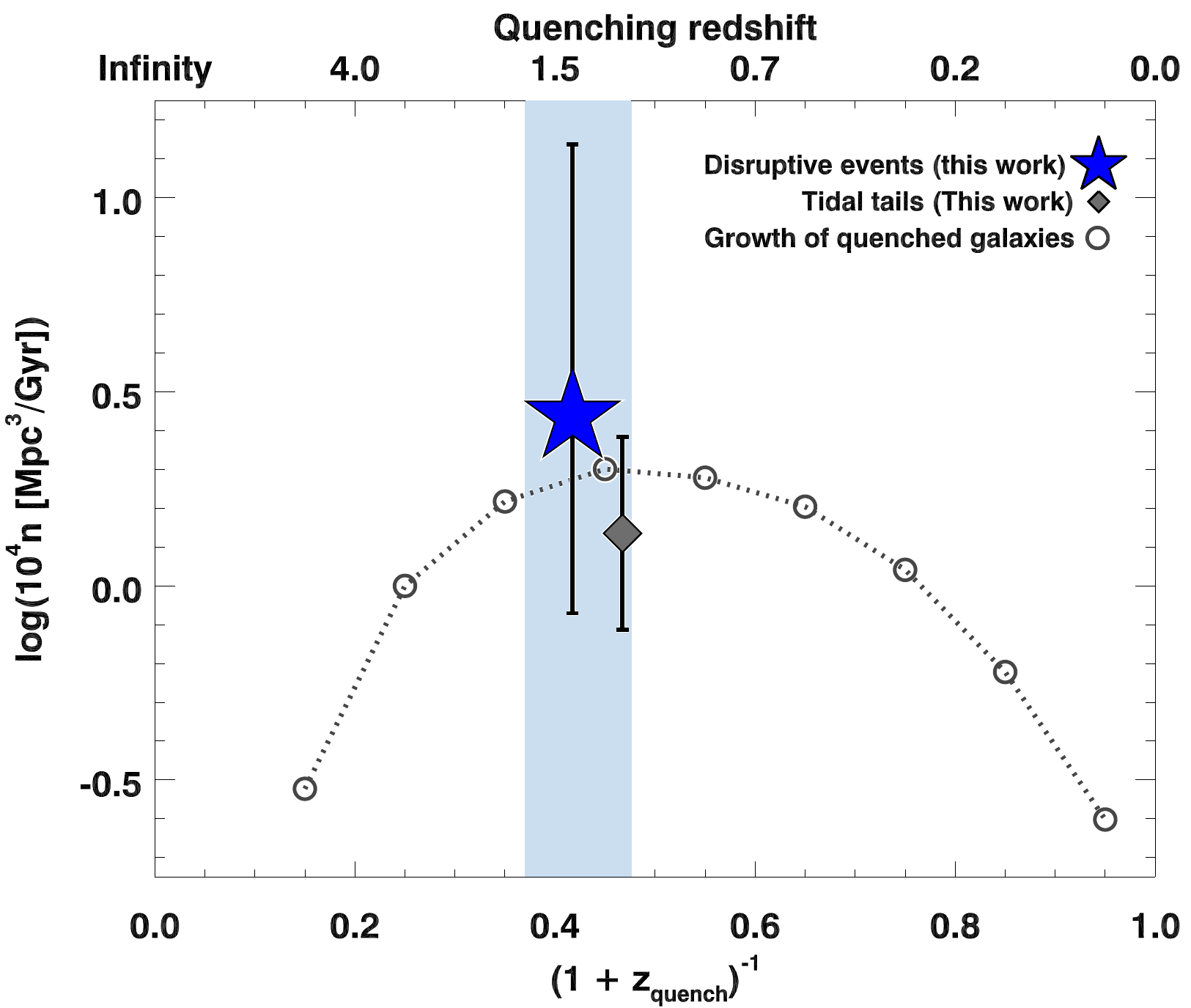}
\caption{ {\bf Comparison between disruptive events rates and density of newly quenched galaxies.}
The blue star represents the rate of disruptive events and associated 1$\sigma$ uncertainty expected in the $1.1 \leqslant z \leqslant 1.7$ redshift range probed by our survey (light blue area). 
The black open circles show the differential number density function inferred from the number density of quiescent galaxies selected above log(M$_{\star}$/M$_{\odot}$) = 10.6 in the ZFOURGE survey \cite{Straatman14}. The 1$\sigma$ uncertainty associated to this quantity is $\sim 0.2$ dex, i.e. much lower than the error associated to our statistics. The stellar mass dependence of the quenched galaxies differential evolution is smaller than the measurements errors in the stellar mass regime probed by our study.
The grey filled diamond and the 1$\sigma$ error shows the density of starburst galaxies expected to eject large tails of gas. This measurement is shifted along the {\it x}-axis for clarity.}
\label{Fig5_number_density}
\end{figure}

\begin{figure}
\centering
\includegraphics[scale=0.9]{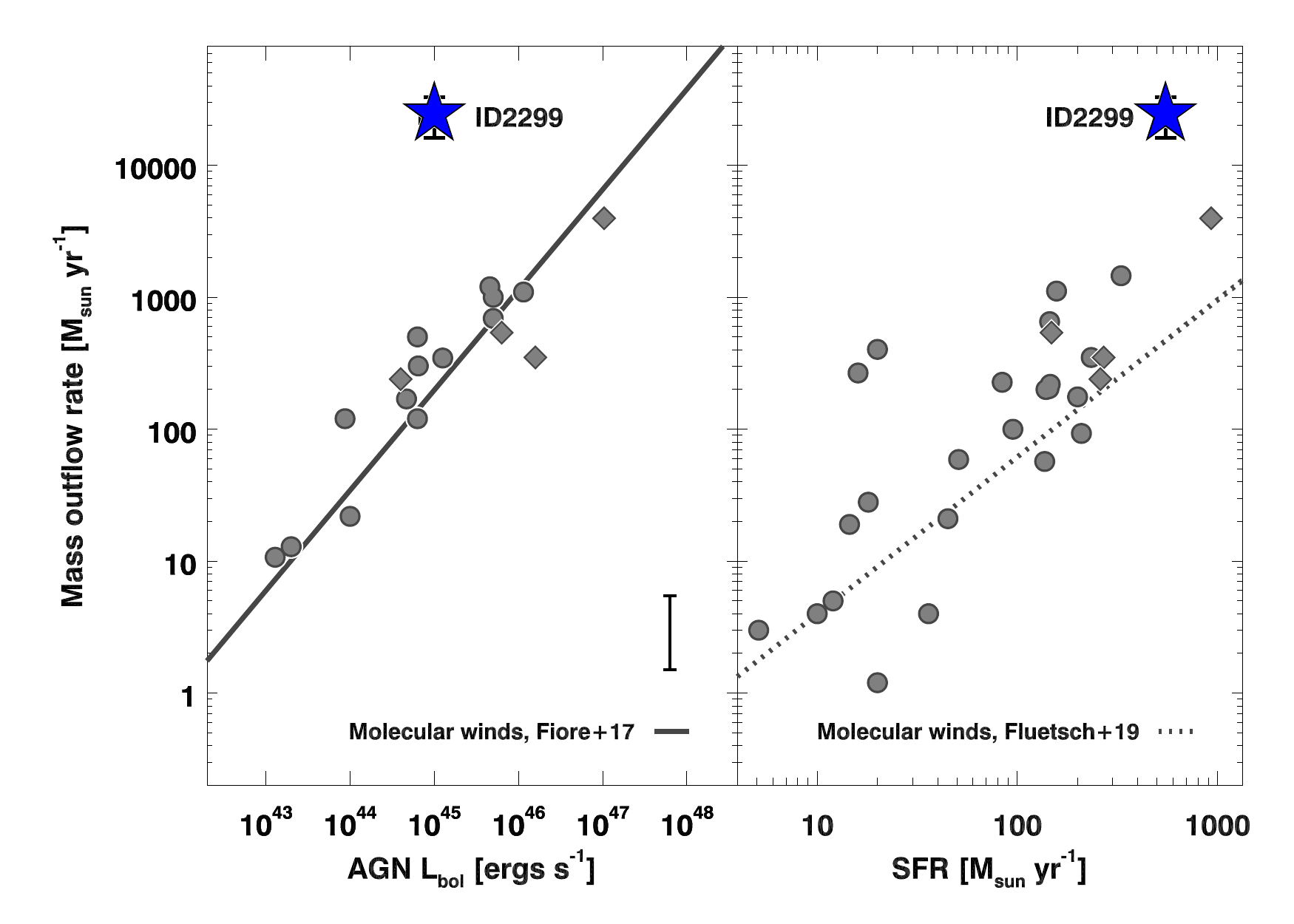}
\caption{ {\bf Comparison between ID2299 and molecular winds from the literature.}
The left panel shows the mass outflow rate as a function of AGN bolometric luminosity. The solid line in this panel shows the scaling relation from Reference \cite{Fiore17}.
The right panel shows the mass outflow rate as a function of SFR. Here, the dotted line displays the scaling relation from Reference \cite{Fluetsch19}.
In both panels, grey filled circles are molecular winds in the local Universe \cite{Fiore17, Fluetsch19} while grey filled diamonds are molecular winds at high redshift \cite{Geach14, Vayner17,HerreraCamus19,Brusa18}.
Some of these measurements have been rescaled to account for the different assumptions in the mass outflow rate computation.
The typical mass outflow rate error for the literature measurements is 0.3 dex (see Ref. \cite{Fluetsch19}) and it is reported in the bottom right corner of the left panel.
The typical uncertainty for the literature measurements of AGN bolometric luminosity and SFR is 0.2 dex and 0.3 dex, respectively \cite{Fluetsch19}.
The mass outflow rate of ID2299 in the feedback-driven wind hypothesis is highlighted with a blue star.
All quoted uncertainties in this figure are at 1$\sigma$ level.}
\label{Fig3_scaling_rels}
\end{figure}

\begin{figure}
\centering
\includegraphics[scale=0.9]{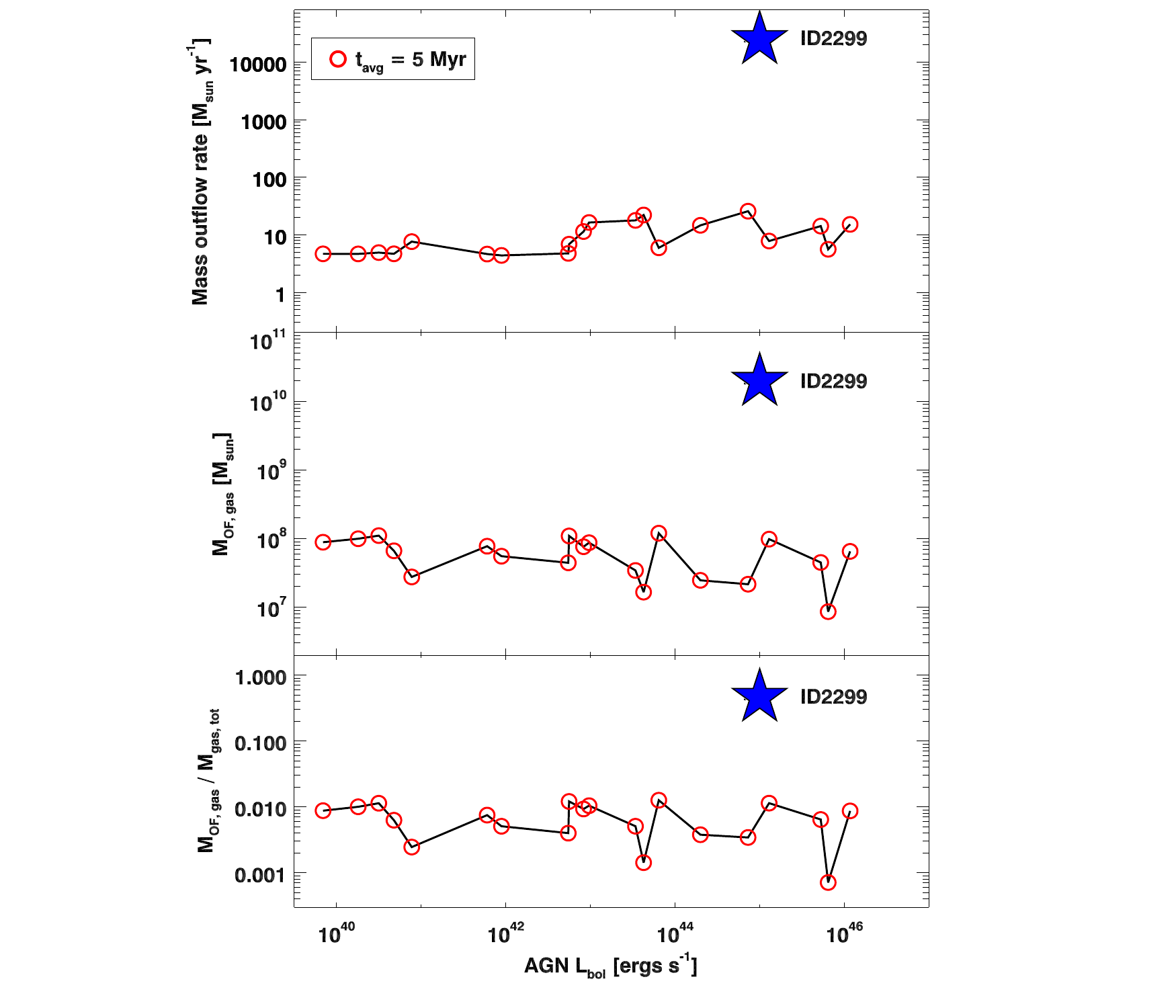}
\caption{ {\bf Comparison between ID2299 and simulations of AGN-driven winds.}
The top panel shows the mass outflow rate as a function of the AGN bolometric luminosity. The middle panel shows the trend for the gas mass in the outflow. The bottom panel shows the trend for the outflowing gas mass normalized by the galaxy gas mass. 
In all panels ID2299 is highlighted with a blue star. 
Error-bars are at 1$\sigma$ level and are smaller than the symbol size.
More details on the simulations are provided in the Methods section.}
\label{Fig4_OF_simulations}
\end{figure}

\begin{figure}
\centering
\includegraphics[scale=0.45]{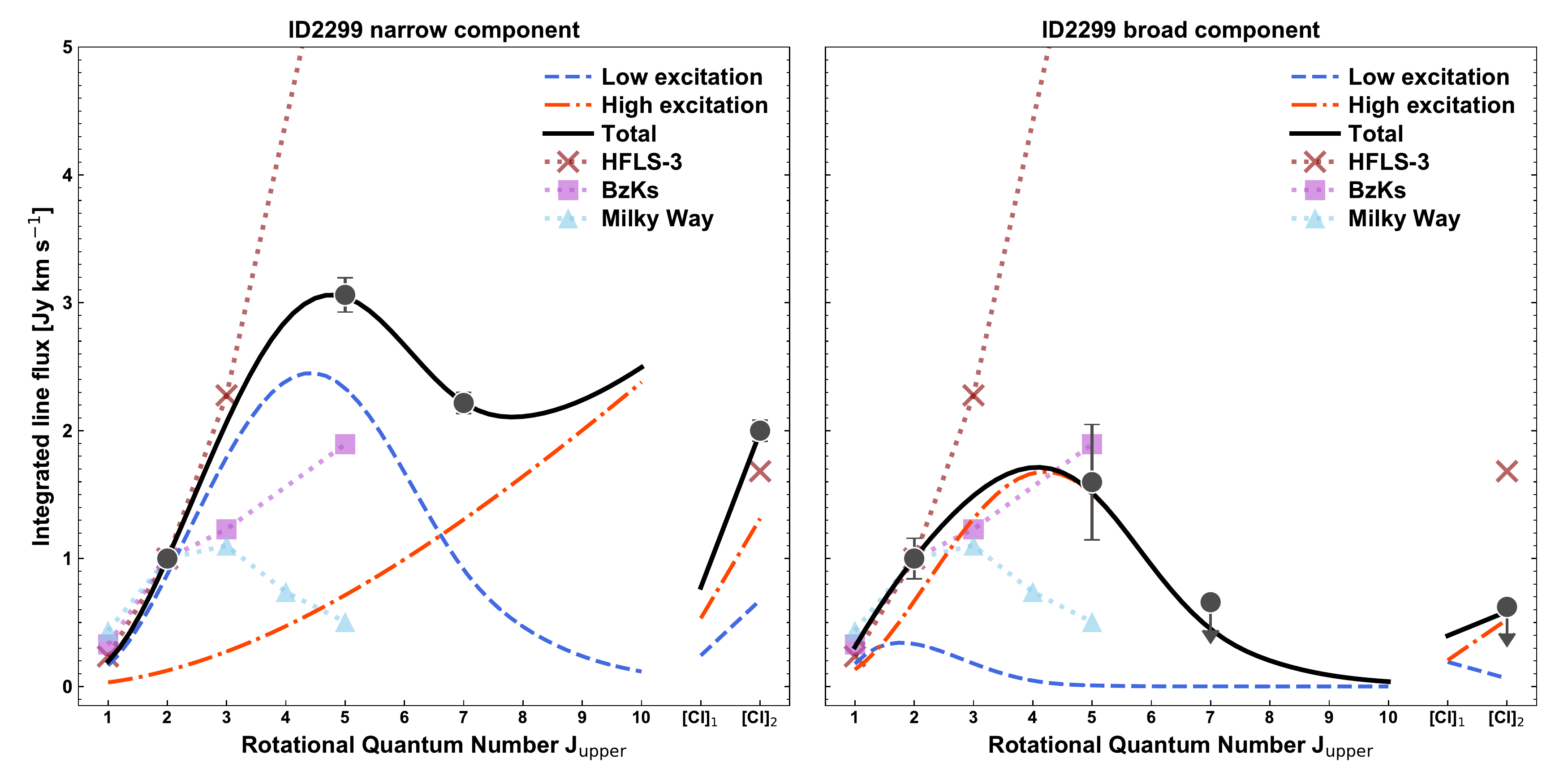}
\caption{  {\bf Molecular gas conditions in the narrow and broad component of ID2299.}
The left and right panel show the best-fit spectral line energy distribution (SLED) from the large velocity gradient (LVG) modelling of the narrow and broad CO/[CI] emission, respectively (see Methods for further details).
Open black circles and 1$\sigma$ errors are the measured data-points. Down-facing black arrows show 3$\sigma$ flux upper limits.
The measurements are normalized to the CO(2-1) flux of each component.
The black solid line is the total best-fit model whereas the dash-dotted red line and the blue dashed line represent the high- and low-excitation LVG models, respectively. 
Additional SLEDs are shown for comparison. }
\label{Fig_LVGmodels}
\end{figure}

\begingroup
\setlength{\tabcolsep}{1pt} 
\renewcommand{\arraystretch}{0.8} 
\begin{table}
\begin{center}
\caption{ {\bf Properties of the galaxy and the ejected material.}
\\$^{\rm a}$ The spectroscopic redshift accuracy is better than $10^{-3}$ (Ref. \cite{Hasinger18})
\label{Table}}
\begin{tabular}{lr}
\hline \hline \noalign {\smallskip}
%Galaxy integrated properties 
$z_{\rm spec}$ & 1.395$^{{\rm a}}$ \\   %specz
M$_{\star}$ &  9.4 $\pm$  1.3 $\times 10^{10}$ M$_{\odot}$ \\ %stellar mass
SFR &  550 $\pm$  10 M$_{\odot}$yr$^{-1}$ \\ %SFR FIR
L$_{\rm AGN}$ & 1.1 $\pm$ 0.3 $\times 10^{45}$ erg s$^{-1}$\\ %Lbol AGN
R$_{\rm eff, ALMA}$ & 1.54 $\pm$ 0.04 kpc\\ %ALMA effective radius
R$_{\rm eff, K_{\rm s}}$ & 3.3 $\pm$ 0.7 kpc\\ %Ks-band effective radius 
%Galaxy properties from ALMA spectra 
v$_{\rm FWHM, mol, n}$ & 120 $\pm$ 13  km s$^{-1}$\\ %FWHM narrow component
I$_{\rm CO(2-1), n}$ & 0.97  $\pm$ 0.06 Jy km s$^{-1}$  \\ %I_CO21 narrow - fluxes from MPFIT, error from SNR
I$_{\rm CO(5-4), n}$ & 2.97 $\pm$ 0.13 Jy km s$^{-1}$ \\ %I_CO54 narrow - fluxes from MPFIT tied, error from SNR
I$_{\rm [CI](2-1), n}$ & 1.94  $\pm$ 0.08 Jy km s$^{-1}$  \\ %I_CI narrow - fluxes from MPFIT, error from SNR
I$_{\rm CO(7-6), n}$ & 2.15 $\pm$ 0.08 Jy km s$^{-1}$ \\ %I_CO76 narrow - fluxes from MPFIT tied, error from SNR
M$_{\rm mol, n}$&  2.3 $\pm$  0.2 $\times 10^{10}$ M$_{\odot}$ \\ %Galaxy molecular gas mass
%Molecular outflow properties from ALMA spectra
v$_{\rm mol, b}$ & 179  $\pm$ 78  km s$^{-1}$\\ %Centroid separation in the velocity space
v$_{\rm FWHM, mol, b}$ & 535  $\pm$ 135 km s$^{-1}$\\ %FWHM broad component
I$_{\rm CO(2-1), b}$ & 0.82  $\pm$ 0.13 Jy km s$^{-1}$  \\ %I_CO21 broad - fluxes from MPFIT, error from SNR
I$_{\rm CO(5-4), b}$ & 1.31  $\pm$ 0.37 Jy km s$^{-1}$ \\ %I_CO54 broad - fluxes from MPFIT tied, error from SNR
I$_{\rm [CI](2-1), b}$ & 0.46  $\pm$ 0.17 Jy km s$^{-1}$  \\ %I_CI broad - fluxes from MPFIT, error from SNR
I$_{\rm CO(7-6), b}$ & $\leqslant$ 0.33 Jy km s$^{-1}$ \\ %I_CO76 broad - fluxes from MPFIT tied, error from SNR
M$_{\rm mol, b}$&  2.0 $\pm$  0.5 $\times 10^{10}$ M$_{\odot}$ \\ %Outflow molecular gas mass
%Ionized outflow properties from DEIMOS spectrum
v$_{\rm FWHM, ion, n}$ & 144 $\pm$ 28  km s$^{-1}$\\ %FWHM narrow component
v$_{\rm ion, b}$ & 92  $\pm$ 47  km s$^{-1}$\\ %Centroid separation in the velocity space
v$_{\rm FWHM, ion, b}$ & 537  $\pm$ 106 km s$^{-1}$\\ %FWHM broad component
L$_{\rm [OII], b}$ & 4.9  $\pm$ 1.5  $\times 10^{42}$ erg s$^{-1}$\\
M$_{\rm ion, b}$&  1.4  $\pm$  0.4 $\times 10^{7}$ M$_{\odot}$\\
\hline \hline \noalign {\smallskip}
\end{tabular}
\end{center}
\end{table}
\endgroup

\newpage

\section*{Methods}

\subsection{\underline{Target information}}
\label{Sect_infos}

{ The source ID2299 at $10^{\rm h} 2^{\rm m} 15.4^{\rm s} \ 1^{\rm o} 54^{\rm '} 5.3^{\rm ''}$} is a galaxy in the COSMOS field located at z = 1.395, a cosmic epoch close to the peak of star-forming and AGN activity of the Universe \cite{MadauDickinson14}. 
ID2299 classifies as a starburst galaxy as its SFR is more than $\sim 5 \times$ higher than the average star-forming galaxies population at similar stellar mass and redshift\cite{Schreiber15}. 

We observed ID2299 as part of a ALMA survey targeting the CO(5-4) transition and $\lambda_{\rm obs} \sim 1.3$ mm underlying continuum in 123 far-IR selected galaxies with $1.1 \leqslant z \leqslant 1.7$ and $L_{\rm IR} \geqslant 10^{12} \ L_{\odot}$. These observations were carried in Band 6 with an average circularized beam of $\sim 0.7 "$ (Program-ID 2015.1.00260.S, PI Daddi). 
The CO(2-1) observations for this source were obtained from a follow-up of this survey targeting this transition in 75 of the 123 galaxies detected at high significance in the main program. The follow-up observations were carried in ALMA Band 3 with an average circularized beam of $\sim 1.5 "$ (Program-ID 2016.1.00171.S, PI Daddi).    
The CO(7-6) and [CI](2-1) observations were acquired from a follow-up targeting 15 of the 123 galaxies in the main program\cite{Valentino19}.
These observations were carried in ALMA Band 7 to detect the two transitions simultaneously with an average circularized beam of $\sim 0.9 "$ (Program-ID 2019.1.01702.S, PI Valentino). 
The ALMA CO(5-4) and CO(2-1) observations and the data reduction procedure are described in a previous paper \cite{Puglisi19}. 
For a detailed description of the full data-set we refer the reader to Reference \cite{Valentino20}.

The rest-frame optical spectrum of ID2299 covering the ${\rm [OII]_{3726, 3729}}$ doublet  was obtained as part of the DEIMOS 10K Spectroscopic survey \cite{Hasinger18}. We retrieved this spectrum from the COSMOS repository ({\url{http://cosmos.astro.caltech.edu/}}).

{  The source has SFR$ = 550 \pm 10$ M$_{\odot}$yr$^{-1} $. This is measured from the bolometric far-IR luminosity of the dust L$_{\rm FIR}$ integrated over $\lambda \in [8 - 1000] \ \mu$m cleaned by the AGN contribution. 
We measure L$_{\rm FIR}$ by fitting the SuperDeblended far-IR/(sub)-millimetre catalog in COSMOS\cite{Jin18} using the procedure described in a previous paper\cite{Jin18}.
We measured the SFR from L$_{\rm FIR}$ using a widely adopted conversion\cite{Kennicutt98} divided by a factor 1.7 to match the initial mass function (IMF) adopted in this work\cite{Chabrier03}. }
The M$_{\star}$ of the object is M$_{\star} = 9.4 \pm 1.3  \times 10^{10}$ M$_{\odot}$ for the same IMF.
The bolometric luminosity of the AGN embedded in the system is L$_{\rm AGN} = 1.1 \pm 0.3 \times 10^{45} \ {\rm erg s^{-1}}$ . 
We derive these quantities through multi-wavelength spectral energy distribution (SED) fitting to the COSMOS broad-band photometry from the UV to the far-IR \cite{Laigle16, Jin18} using the CIGALE code \cite{Noll09} accounting for the AGN contribution as described in a previous paper\cite{Circosta18}. 
The best-fit SED of the source is shown Extended Data Figure 1.

\subsection{ \underline{Merger classification}}

{  
ID2299 is undergoing a merger. 
This is suggested by its intense star-forming activity as several works show the connection between starburst and merging activity\cite{Silverman15, Silverman18, Puglisi17, Cibinel19, Calabro18, Calabro19}. This intense star-forming activity is hosted in a compact star-forming core of R$_{\rm eff, ALMA} \sim 1.54$ kpc. Such compact star-forming regions can only arise from mergers\cite{Puglisi19}. 
Furthermore, ID2299 is heavily dust obscured. We measure { A$_{\rm 1600} = 6 \pm 0.2$ mag} from the UV-slope $\beta$\cite{Meurer99} and 
{ SFR$_{\rm FIR}$/SFR$_{\rm UV, obs} = 500 \pm 50$}. The large dust attenuation measured for ID2299 is another clear indication of merging activity\cite{Calabro18}.

Several quantitative criteria can be used to classify ID2299 as a merger, analysing available HST F814W image (see Figure \ref{Fig2_maps}).
ID2299 visually classifies as a merger with an interaction phase of four (class MIV) \cite{Kartaltepe10}.
This corresponds to a post-coalescence phase in which tidal tails and disturbed morphology are still clearly visible and the AGN is still active (see Fig. 2 in Reference \cite{Kartaltepe10}).
We measure on the HST F814W image (see Fig.~\ref{Fig2_maps}) a Shape asymmetry index \cite{Pawlik16} of 0.28 which is larger than the minimum value of 0.2 required to classify a galaxy as a merger according to Reference \cite{Pawlik16}.
We further measure a Gini and M20 parameter of 0.51 and -0.83 respectively classifying ID2299 as a merger using the method detailed in Reference \cite{Lotz08}.}

\subsection{ \underline{Size measurement}}

{  We consider the full ALMA data-set discussed in this work to measure the source size in the sub-millimetre following the procedure described in a previous paper\cite{Puglisi19}. 
This procedure uses GILDAS-based (\url{http://www.iram.fr/IRAMFR/GILDAS}) scripts to extract the signal amplitude as a function of the \textit{uv} distance from each tracer. 
The signal for each tracer is combined by scaling the \textit{uv}-distances to a common frequency and marginalizing over a free normalization constant.
The size and its $1 \sigma$ uncertainty are determined by comparing the \textit{uv} distance vs. amplitude distribution to circular Gaussian models. 
The galaxy best-fit size is defined as the effective radius R$_{\rm eff}$ = FWHM/2. 
The goodness of the fit is estimated from the $\chi^2$ minimization as a function of the size. The scripts compare the best-fit $\chi^2$ to the $\chi^2$ for a point source to quantify the probability of the galaxy to be unresolved, $P_{\rm unres}$. A source is considered to be resolved when P$_{\rm unres} \leqslant 10 \%$. 
We show in Extended Data Figure 2 the ID2299 best-fit model in the {\it uv} plane compared to the {\it uv} distance vs. amplitude distribution for the various tracers available. The Figure shows that the various tracers used to estimate the size are similarly extended, as expected from previous results\cite{Puglisi19}. The galaxy best-fit size obtained by combining the full data-set is R$_{\rm eff, ALMA} = 1.54 \pm 0.04 $ kpc with P$_{\rm unres} = 0 \%$. This is consistent within 2$\sigma$ with the size measurement based on the sole CO(2-1) and CO(5-4) data-set previously reported\cite{Puglisi19}.}

\subsection{ \underline{Spectral analysis}}
\label{Spec}

{  We fit the spectra shown in Figure \ref{Fig1_spectra1d} using scripts based on the IDL routine \textsc{mpfit}\cite{Markwardt09}. 
We quote in the following the emission line parameters and their 1$\sigma$ uncertainties as derived from \textsc{mpfit}.

The ALMA CO(2-1), CO(5-4) and [CI](2-1)+CO(7-6) spectra are extracted fitting the data in the UV space accounting for source extension modeled as a Gaussian  of radius R$_{\rm eff, ALMA} = 1.54$ kpc corresponding to the effective radius of the galaxy in the sub-millimeter\cite{Puglisi19}.
We fit and subtract the continuum using a power law in the form S$_{\nu} =$ S$_{0} \times (\frac{\nu}{\nu_{0}})^{3.7}$. The normalization is $\nu_{0} = 101.25 \ {\rm GHz}$, $\nu_{0} = 248.40  \ {\rm GHz}$ and $\nu_{0} = 342.55 \ {\rm GHz}$ for the CO(2-1), CO(5-4) and [CI](2-1)+CO(7-6) spectra, respectively. 
The continuum underlying the CO(2-1) emission is almost undetected (SNR $\sim$ 2, S$_{0} = 0.05 \pm 0.02 \ {\rm mJy}$). 
Instead, we detect a strong continuum under the CO(5-4) emission (SNR $\sim$ 14, S$_0 = 1.80 \pm 0.12  \ {\rm mJy}$) and below the [CI](2-1)+CO(7-6) emission (SNR $\gtrsim 100$, S$_{0} = 5.04 \pm 0.15 \ {\rm mJy}$).
We use Gaussian functions to model the emission lines in the continuum-subtracted spectra leaving all the fit parameters free to vary. 
The resulting emission line centroid and velocity dispersion measured in each spectrum are remarkably consistent. 
However, the single Gaussian fits yield significant residuals suggesting that an additional component is required (light blue curve in the bottom panels of Fig.~\ref{Fig1_spectra1d}, see also the next section).  
Therefore, we fit two Gaussian functions to the CO(2-1) spectrum leaving all the fit parameters free to vary. 
We detect an additional component red-shifted by 179 $\pm$ 78 kms$^{-1}$ from the narrow component centroid. The FWHM of this component is v$_{\rm FWHM, b, CO(2-1)} = 535 \pm 135$ kms$^{-1}$. 
We thus fit two Gaussian functions also to the CO(5-4) and [CI](2-1)+CO(7-6) spectra tying the centroid position and line width to the line parameters measured at higher significance in CO(2-1). 
We detect a broad CO(5-4) emission at 3.5$\sigma$ and a [CI](2-1) broad emission at 2.7$\sigma$. 
We derive a 1$\sigma$ upper limit for the broad CO(7-6) emission (Table \ref{Table}).}

We model the continuum in the DEIMOS spectrum using a constant function. This has a normalization F$_{\rm cont} = 0.638 \pm 0.02 \times \ 10^{-17}$ erg/s/cm$^2$/{\AA} as measured from an emission lines free region close to the ${\rm [OII]_{3726, 3729}}$ doublet.
We fit two narrow Gaussian to the ${\rm [OII]_{3726, 3729}}$ doublet in the continuum-subtracted spectrum. We fix the wavelength separation to the theoretical value imposing the FWHM to be identical. 
Fitting a narrow Gaussian doublet yields residuals at $\gtrsim 15 \sigma$ significance (Figure \ref{Fig1_spectra1d}).
{ We add two broad Gaussian function to the fit, fixing their wavelength separation and imposing an identical FWHM.
We fix the flux ratio of the narrow and broad Gaussian functions to the theoretical values in the low-density limit ${\rm I_{[OII]_{3729}}/I_{[OII]_{ 3726}} = 1.5}$.}
We measure for the broad components of the doublet a { 92 $\pm$ 47 kms$^{-1}$} shift from the narrow centroids. 
This is smaller than the velocity shift measured in the ALMA spectra.
This difference might be associated to the larger uncertainties in fitting the ${\rm [OII]_{3726, 3729}}$ doublet.
However, the measurements are consistent within the 1$\sigma$ uncertainties. 
The broad components detected in the ${\rm [OII]_{3726, 3729}}$ have { v$_{\rm FWHM, b, [OII]} = 537 \pm 106$ kms$^{-1}$}.

\subsection{ \underline{Significance of the broad component detection in the sub-millimetre}}
\label{Sect_significance}

We fit the CO(2-1) spectrum with a single Gaussian component having v$_{\rm FWHM, n}$ and redshift consistent with measurements performed on the [OII], CO(7-6), [CI](2-1), and CO(5-4) spectra. 
We scan the residual finding positive residual emission over 11 channels with integrated SNR$=4.5$.
We estimate the likelihood that such feature could arise from noise following a method that requires to estimate the number of effective independent trials in the scan\cite{Jin19}.
The number of independent trials in our spectral scan is of order of 47, considering the width of the feature and the velocity range of our search. 
This implies a probability of $1.7\times10^{-4}$ that the detected feature arises from noise. 
We apply this approach also to the residuals obtained after fitting the CO(2-1) spectrum with two components.
The strongest feature detected has an integrated SNR$=2.6$ and extends over 5 channels with a { $-1500$ kms$^{-1}$ velocity shift from the narrow component centroid}.
This feature has a 0.96 probability to arise from noise fluctuation. 
We thus conclude that the CO(2-1) spectrum requires two components with high confidence (spurious probability at the level of $1.7\times10^{-4}$).
The overall SNR of the broad component in the other ALMA spectra is not high enough to perform an independent search. 

Having established that the CO(2-1) observed spectrum requires two components in the fit, we perform a narrow+broad Gaussian components fit in the observed spectrum as described in the previous section. We subtract the narrow component and we apply the method described above searching for the most significant feature in the residuals. This leads to a detection of a CO(2-1) line with an integrated SNR$=6.2$ over 14 velocity channels. 

We note that the broad component parameters derived from the CO(2-1) fit are consistent with those of the [OII] emission. 
The chances of obtaining a spurious multiple detection with consistent kinematics are basically zero.

\subsection{ \underline{Imaging the broad component in the ALMA cubes}}
\label{Sect_images}

To determine the spatial centroid and the physical extension of the broad emission while minimizing the contamination from the stronger galaxy emission we subtract the narrow components from the CO(2-1), CO(5-4) and [CI](2-1)+CO(7-6) data-cubes, following the procedure described below.

We perform the subtraction in the \textit{uv} plane using a suite of automated scripts within GILDAS.
We use the best-fit model derived by fitting the one-dimensional spectrum (Figure \ref{Fig1_spectra1d}) to create a model for the continuum and narrow line flux at each frequency. 
We use the \textsc{uvfit} task in GILDAS to subtract this model at each frequency within the effective radius of the galaxy molecular gas emission (Table \ref{Table}). We iterate the procedure over the full frequency range of the spectral window in which the line is detected. 
We then recombine the residuals obtained at each step to reconstruct the cube containing the broad component emission clean from the galaxy contribution.
The spectrum extracted from this data-cube within an ``optimal'' aperture (i.e. the aperture that maximizes the SNR of the emission) is fully consistent with the spectrum extracted over the galaxy in the original cube shown in Figure \ref{Fig1_spectra1d}.
We note that we cannot estimate a physical size for the broad emission because of the low SNR of the detection. However, the size over which the broad component spectrum is extracted is consistent with the narrow component size although with large uncertainties given the low SNR of the broad component detection.

We show in Figure \ref{Fig2_maps} the broad component line images obtained from these cubes .
These cubes show significant emission over the spatial extent of the galaxy. The centroid of the broad emission has no significant off-set with respect to the galaxy. 
{We note that the line images of the CO(5-4), [CI](2-1) and CO(7-6) broad components show high levels of noise as expected from the SNR of the detection in these transitions.}

\subsection{ \underline{Mass outflow rate computation}}
\label{Sect_MOFR}

In the assumption that the broad component is associated to a feedback-driven wind, we compute the mass outflow rate following the prescription adopted to compute the scaling relation in the left panel of Figure~\ref{Fig3_scaling_rels}\cite{Fiore17}.
This is to minimize the effects of geometry assumptions and systematics when comparing with the literature results. Our choice thus ensures a robust determination of differential quantities in the left panel of Figure \ref{Fig3_scaling_rels}. 
We note that the scaling relation in the right panel of Figure \ref{Fig3_scaling_rels} uses a slightly different recipe for the mass outflow rate computation\cite{Fluetsch19}. This results in mass outflow rates underestimates by a factor $\sim 0.5 \ dex$ on average. Considering this correction factor to account for different assumptions in the mass outflow rate computation does not alter the conclusion of this work.

We thus compute the mass outflow rate as:
\begin{equation}
\dot{M}_{\rm OF} = \frac{3 \times v_{\rm max} \times M_{\rm gas}}{R_{\rm OF}}
\label{Eqn_mofrate}
\end{equation}

where R$_{\rm OF}$ is the radius at which the outflow rate is computed and v$_{\rm max}$ is the wind maximum projected velocity. Consistently with the literature definition\cite{Fiore17}, we measure v$_{\rm max}$ as:

\begin{equation}
v_{\rm max} = (v_{\rm broad} - v_{\rm narrow}) + 2\times\sigma_{\rm broad} = 634 \pm 139 \ {\rm km s^{-1}}
\end{equation}

where v$_{\rm broad}$ and v$_{\rm narrow}$ are velocity peaks of the broad and narrow components respectively and $\sigma_{\rm broad}$ is the standard deviation of the broad component emission from the ALMA spectra.

We derive the outflowing gas mass from the integrated CO(2-1) luminosity of the broad component log(L$^{'}_{\rm CO(2-1), broad}/({\rm K km^{-1} pc^2})$) = 10.32. 
This is derived from the integrated flux reported in Table \ref{Table} following eqn. 2 in Reference\cite{Silverman18b}.
We then convert this quantity to L$^{'}_{\rm CO(1-0)}$ by assuming starburst-like CO excitation (L$^{'}_{\rm CO(2-1)}$/L$^{'}_{\rm CO(1-0)} = 0.85$\cite{Silverman18b}). Finally we compute the gas mass as M$_{\rm mol} = \alpha_{\rm CO} \times {\rm L}^{'}_{\rm CO(1-0)}$ by adopting a ULIRG-like conversion factor $\alpha_{\rm CO} = 0.8 \ {\rm M}_{\odot}{\rm (K km s^{-1} pc^2)}^{-1}$, as described in the main text. This is consistent with literature assumptions\cite{Fiore17} and it is also justified by the ISM conditions of the host galaxy.
We note however that the CO SLED of the outflow (Fig.~\ref{Fig_LVGmodels}) implies a less excited material suggesting  $\alpha_{\rm CO} = 3.6 \ {\rm M}_{\odot}{\rm (K km s^{-1} pc^2)}^{-1}$. The use of a larger conversion factor would increase the gas mass (and gas fraction) carried in the outflow. 
Instead, from the CO SLED of the broad component we extrapolate  L$^{'}_{\rm CO(2-1),b}$/L$^{'}_{\rm CO(1-0),b} = 0.81$ which is consistent with the value of 0.85 adopted above. 

We finally compute $\dot{\rm M}_{\rm OF}$ at the effective radius of the molecular gas emission as measured on the ALMA maps. We thus define R$_{\rm OF} =$ R$_{\rm eff, ALMA} = 1.54 \ {\rm kpc}$. This follows by construction as the one-dimensional spectrum, from which the outflow parameters are derived, is extracted over a beam representing the galaxy  extension. Furthermore, as discussed in the previous section of the Methods and shown in Figure \ref{Fig2_maps}, there is no evidence for the outflow being more extended than the galaxy. 

We derive errorbars for the mass outflow rate computation by propagating the 1$\sigma$ uncertainties over the molecular gas mass, the maximum velocity and the outflow radius.

For a complete energy budget of the outflow, we should account for all the phases of the gas in which the outflow is detected. Here we detect the outflow also in the ${\rm [OII]_{3726, 3729}}$ spectrum sampling the ionized phase of the phenomenon. 
Following Eqn. 5 in Reference \cite{Carniani15}, we measure the ionized gas mass in the outflow M$_{\rm OF, ion}$ from the ${\rm [OII]_{3726, 3729}}$ broad component luminosity (Table \ref{Table}).
{ This computation requires to measure the electron density in the outflowing material\cite{Carniani15} but this quantity is poorly constrained by our data. 
As such, we assume that the outflowing gas has the same electron density as the galaxy and we assume n$_{\rm e} = 300 \ {\rm cm^{-3}}$ for the latter, corresponding to the typical electron density of star-forming galaxies at $z \geqslant 1$\cite{ForsterSchreiber19}.
We use a [OII] emissivity j$_{\rm [OII]} = 1.641 \times 10^{-21} \ {\rm erg s^{-1} cm^{-3}}$ as derived with the \textsc{temden} package in IRAF at T$_{\rm e} = 10^4 \ {\rm K}$ and n$_{\rm e} = 300 \ {\rm cm^{-3}}$.
We do not apply a dust attenuation correction to the observed [OII] luminosity and we assume solar metallicity considering that starburst galaxies at high-redshift are metal-rich \cite{Puglisi17}. 
We finally obtain that the ionized gas mass in the outflow is M$_{\rm OF, ion} = 1.4 \pm 0.4 \times 10^7$ M$_{\odot} $. 
To derive the mass outflow rate in the ionized phase we use Eqn. \ref{Eqn_mofrate} considering the outflowing gas mass in the ionized phase M$_{\rm OF, ion}$. This corresponds to a mass outflow rate $\dot{\rm M}_{\rm OF, ion} = \ 20 \pm$ 7 M$_{\odot}$yr$^{-1}$}. 
The ionized gas mass and mass outflow rate are negligible in the mass/energy budget of the outflow and the molecular phase of the gas dominates the energetic of the phenomenon, carrying $\geqslant \times 1200$ more energy than the warm ionized phase.
This is somewhat expected from the few multi-phase studies of outflows at high redshift although typical molecular-to-ionized outflow rates ratios do not seem to exceed a factor of $\sim 10$ (Ref. \cite{HerreraCamus19}).
Since the molecular gas substantially dominates the mass outflow rate and because the physical quantities associated to the ionized gas are affected by large uncertainties due to metallicity, electron density and dust attenuation assumptions, we refer only to the molecular phase of the gas through our discussion in the main text.

In the tidal ejection scenario, the broadening and shift of the line are produced by a combination of large scale and internal velocity gradients as well as complex projection effects\cite{Bournaud04}. This implies that using $v_{\rm max}$ in Equation \ref{Eqn_mofrate} may not be representative of the velocity of the ejected gas.
We further note that Equation \ref{Eqn_mofrate} assumes that a conical outflow is continuously refilled with ejected clouds \cite{Cicone14}.
A tidal tail is instead a single event ejection but this is yet different from a "shell-like geometry" in a single explosive ejection from a feedback-driven wind.
The neutral gas morphology of local mergers \cite{HibbardYun96} suggest in fact that the ejected gas is more uniformly distributed along the tail. 
We thus conclude that, given the lack of constraints on the gas geometry, assuming a simple scenario where the ejected gas is uniformly distributed within a conical volume is a valid approximation for the mass outflow rate also in the tidal ejection scenario.
Using Equation \ref{Eqn_mofrate} and conservatively assuming that the average velocity shift of the broad component is representative of the velocity of the ejected gas, we obtain a mass outflow rate of $\dot{\rm M}_{\rm mol, OF} = 7100 \ \pm \ 3600 $ M$_{\odot}{\rm yr}^{-1}$ for the tidal ejection scenario. 

\subsection{\underline{Simulations of feedback-driven outflows}}

{ 
We compare our results to state-of-the-art hydrodynamical simulations of AGN-driven winds inside gas-rich massive galaxies presented elsewhere\cite{GaborBournaud14}. 
These simulations include models for cooling, star formation, black hole growth, supernova and AGN feedback.
The central black hole in the simulated galaxies undergoes a stochastic growth via high accretion episodes lasting $\sim$ 10 Myr at a time.
The aim of these simulations is to study the effects of AGN feedback in typical galaxies of the high-redshift Universe. 
As such, the simulated galaxies are massive ($10^{10} \leqslant$ M$_{\star} \leqslant 10^{11}$ M$_{\odot}$) and gas-rich (gas fractions above $\sim 50\%$). 

These simulations provide us trends for the mass outflow rate, the black hole accretion rate and the galaxy dense gas mass (Figure 3 and Figure 7 in Reference\cite{GaborBournaud14}).
We integrate these trends on 5 Myr time steps to quantify the mass outflow rate and outflowing gas mass as a function of the AGN bolometric luminosity shown in Figure \ref{Fig4_OF_simulations}. 
We convert the black hole accretion rate to an AGN bolometric luminosity using a 10$\%$ accretion efficiency.
To measure the gas fraction in the simulated galaxy (bottom panel of Figure \ref{Fig4_OF_simulations}) we normalize the outflowing gas mass to M$_{\rm gas, gal} = 1.3 \times 10^{10}$ M$_{\rm \odot}$ which is the gas mass of the simulated galaxy at t = 0 in a nuclear region of 5 kpc (see Figure 7 in Reference \cite{GaborBournaud14}).

We show in Fig.~\ref{Fig4_OF_simulations} that the mass outflow rate, total mass expelled, and fraction of ISM mass expelled from the parent galaxy predicted by the simulations are all at least two orders of magnitude higher in ID2299 than what AGN-feedback can produce, regardless of the underlying current or past AGN luminosities. 
These conclusions are in line with other works, showing that the impact of AGN on the star formation history of the host galaxy is marginal \cite{Roos15}. 
This is also consistent with recent studies demonstrating that the molecular gas associated to feedback-driven outflows is scarce\cite{BiernackiTeyssier18}.
Alternative models considering efficient gas cooling in the outflow can form few $\sim 10^9$ M$_{\rm gas}$ of outflowing molecular gas \cite{Costa14, Costa15} and this is one order of magnitude lower than our observed values. 
Models reproducing extreme mass outflow rates fall short by 1 order of magnitude to reproduce the outflowing molecular gas mass in ID2299 while predicting substantially larger outflow velocities\cite{Costa18}.
Finally, models considering the formation of molecular gas within the outflow significantly underpredict the outflow velocities and outflowing molecular gas by two orders of magnitude \cite{RichingsFS18a,RichingsFS18b}}

\subsection{ \underline{Excluding inflow scenarios}}

{The galaxy has a stellar mass M$_{\star} \sim 10^{11}$ M$_{\odot}$ (Table \ref{Table}). 
The corresponding hosting dark matter halo is thus expected to be on average M$_{\rm halo} \sim 10^{13}$ M$_{\odot}$\cite{Behroozi13}.
Such halo is not expected to have cold accretion at $z = 1.4$\cite{Dekel09}.
Even allowing for a substantial scatter in the stellar-to-halo mass ratio, and assume that the hosting halo mass is M$_{\rm halo} \sim 10^{12}$ M$_{\odot}$ we would expect cold accretion rates of $\dot{\rm M}_{\rm inf, gas} \sim (10 - 50)$ M$_{\odot}$yr$^{-1}$ spread over a $\sim 50$ kpc scale \cite{Goerdt10, Dekel13}. 
This inflow rate is several orders of magnitude lower than the mass outflow rate we measure over $\sim \times 10$ smaller scales.
It thus seems highly implausible that the broad emission is associated to gas infalling from the cosmic web.

Similarly, assuming that this is a fountain-like event with previously ejected gas returning to the galaxy would appear implausible when considering that about half of the ISM of the system is present in the extended gas and at small spatial scales, fountain-like material would be diffuse. }

\subsection{ \underline{Rejecting the ``fossil outflow'' interpretation}}

Following the scaling relation in the left panel of Figure \ref{Fig3_scaling_rels}\cite{Fiore17} the AGN must have been as bright as L$_{\rm bol} \sim 10^{47.8} \ {\rm erg\ s^{-1}}$ to trigger the observed disruptive event.
This corresponds to a X-ray AGN luminosity L$_{\rm X-ray} \gtrsim\ 10^{46} \ {\rm erg\ s^{-1}}$ (see Reference \cite{Lusso12}). 
Such a luminous quasar would likely display traces of radio AGN activity \cite{Merloni03} having a variability time-scale substantially longer than those observed in the X-rays ($\gtrsim$ 100 Myr against $\lesssim$1 Myr, see Ref. \cite{Morganti17}). 
However, the 1.4-3 GHz radio luminosity of ID2299 is fully accounted by star formation \cite{Delvecchio17}.
Deep VLBA 1.4 GHz observations\cite{HerreraRuiz17} at $0.01''$ resolution sampling circumnuclear scales (85 pc at z=1.4) where the core AGN jet is supposed to be launched, give a 3$\sigma$ upper limit log(L$_{\rm radio}/{\rm W\ Hz^{-1}}) \leq 23.55$. The probability that this radio luminosity is associated to a luminous X-ray quasar is less than 35\% (see Reference \cite{Lafranca10}).
The lack of radio AGN signatures at high and low resolution disfavours the possibility that the AGN hosted in ID2299 was as powerful as L$_{\rm bol} \geqslant 10^{47.8}$ erg/s in the past. 
More relevantly, the required AGN bolometric luminosity approaches that of the most luminous known QSO in the Universe\cite{Wolf18}. The space density of similar QSOs is of order of $10^{-9}$~Mpc$^{-3}\;$\cite{Delvecchio20}, many orders of magnitude below those of galaxies in our survey, making it highly implausible that one such monster just switched-off from one of the 123 galaxies in our survey.

\subsection{\underline{Arguments against a two-galaxy merging scenario}}
\label{Sect_merger}

The broad component detected in the ID2299 spectra might be associated to a galaxy  of an unresolved merging system of two galaxies. 
However, we demonstrate in the following that this possibility can be excluded using the available constraints on the dynamical and the stellar mass of the system.

In the assumption that the broad component is a galaxy in a merging pair, we derive its dynamical mass by using a formula which was explicitly calibrated using numerical simulations of merging pairs\cite{Silverman18}:
 
\begin{equation}
M_{\rm dyn, r_{\rm eff}} = f_{\rm c} \times \frac{r_{\rm eff} \times (v_{\rm FWHM}/2)^2}{G {\rm sin}^2 i}
\label{Eqn_mdyn}
\end{equation}

This formula is based on a thin-disk rotator and it includes a correction factor $f_{\rm c}$ to account for physical effects biasing the dynamical mass measurement in a merger. These effects include turbulence, velocity dispersion, clumpiness and inflows. 
The correction factor $f_{\rm c}$ is derived as the difference between the ``observed'' and ``true'' dynamical mass in simulations of gas-rich galaxy mergers\cite{Fensch17} (see also Sect. 5.2 and Fig.~10 in Reference \cite{Silverman18}). 
We use $f_{\rm c} = 10^{0.45} = 2.81$ which is the median ratio of the log$_{\rm 10}({\rm M}_{\rm dyn, obs}/{\rm M}_{\rm dyn, true})$ distribution plotted in Fig. 10 of Reference \cite{Silverman18}. 
For the inclination we use $i = 57^{\circ}$ corresponding to the average inclination angle of randomly orientated galaxies \cite{Coogan18}.
Finally, we compute the total dynamical mass of the galaxy:

\begin{equation}
M_{\rm dyn,tot} = 2 \times M_{\rm dyn, r_{\rm eff}}
\label{Eqn_mdyn_tot}
\end{equation}

This factor of two correction is required because only half of the light/baryonic mass are contained within $r_{\rm eff}$, by definition.
Using Eqn. \ref{Eqn_mdyn} and \ref{Eqn_mdyn_tot} with R$_{\rm eff, ALMA}$ = 1.54 kpc and v$_{\rm FWHM, mol, b}$= 535 km s$^{-1}$ we obtain M$_{\rm dyn, tot, b} = 2.1 \times 10^{11}$ M$_{\odot}$.
The uncertainty associated to M$_{\rm dyn, tot, b}$ is 0.3 {\it dex} as derived from a Montecarlo simulation, accounting for errors in the { correction factor f$_{\rm c}$ (0.2 $dex$\cite{Silverman18})}, the effective radius and the velocity dispersion.
The additional uncertainty associated to the unknown inclination can substantially affect this estimate only to the upper-side { (see Figure 11 in Reference \cite{Silverman18}}).
The dynamical mass obtained through Equations \ref{Eqn_mdyn} and \ref{Eqn_mdyn_tot} represents the total dynamical mass of the system as seen by ALMA. This compares to a gas mass M$_{\rm mol, b} \sim 2 \times 10^{10}$ M$_{\odot}$ over the same spatial scales of $\sim 1.54$ kpc. The dark matter contribution is quite negligible at these very compact scales \cite{Genzel17}.

We use \textsc{galfit} \cite{Peng10} to estimate the K$_{\rm s}$-band effective radius of ID2299 (tracing the stellar mass distribution at $z \sim 1.4$) following the method detailed in a previous paper \cite{Puglisi19}. 
We obtain R$_{\rm eff, K_{s}} = 3.3 \pm 0.7$ kpc and this is $2.2 \pm 0.5 \times$ larger than R$_{\rm eff,ALMA}$.
This implies that the stellar mass extends over an area that is $4.6 \pm 2 \times$ larger than the ALMA one over which the dynamical and gas mass are measured, as seen in a substantial fraction of IR-luminous resolved ALMA galaxies \cite{Tadaki17, Elbaz18, CalistroRivera18, Puglisi19, Franco20}.
{ As such, the dynamical mass derived above does not account for the more extended stellar distribution not associated with the central starburst.}
To account for the different stellar mass and gas distribution in ID2299, we re-scale the total stellar mass of the system to the ALMA area following Reference \cite{Puglisi19}:
\begin{equation}
\rm M_{\rm \star, ALMA} = M_{\star}  \times (Area_{\rm ALMA} / Area_{\rm \star}) = M_{\rm \star}  \times (R_{\rm eff,ALMA} / R_{\rm eff, K_{\rm s}})^2 
\label{Eqn_escaled}
\end{equation}
where M$_{\rm \star}$ is the total stellar mass derived via SED fitting (see Table \ref{Table}).
We obtain M$_{\rm \star, ALMA} = 2.1 \times 10^{10}$ M$_{\odot}$ and this includes the contribution from the main galaxy and the putative merging companion. Conservatively assuming a 50\% splitting between the two putative galaxies and adding to the gas mass, the total falls short by a factor of 0.85 $dex$ ($\sim 7\times$) with respect to M$_{\rm dyn, tot, b}$.
{
We thus conclude that the discrepancy between the dynamical and baryonic mass is significant, also when accounting  for { the reported observational} uncertainties { as well as the systematic uncertainties associated to} the dynamical status of the system \cite{Silverman18}.}
We note that the stellar mass is robust against dust attenuation effects. This is because we include the K$_{\rm s}$ and IRAC photometric bands in the SED fitting (see Extended Data Figure 1), corresponding to 1--3$\mu$m in the rest-frame. The emission at these rest-frame wavelengths is known to get aligned with  CO and radio even for the most obscured starbursts \cite{Tan14, Silverman15, Silverman18, Wang19} and is thus representative of the bulk of the stellar mass. 
{ Barring further systematic effects we might not be aware of, this strongly suggests that the broad emission observed in the ID2299 spectra is associated to unbound outflowing material. This seems to be the most plausible assessment of the mass budget of the source given our current understanding of the system. }

\subsection{ \underline{Molecular gas conditions of the galaxy and of the ejected material}}
\label{Sect_sleds}

We use the flux density of the narrow and broad component emission in the ALMA spectra to construct the  spectral lines energy distribution (SLED) of the galaxy and the expelled material, respectively (Figure \ref{Fig_LVGmodels}). The flux densities are re-normalized to a CO(2-1) flux of unity.
We apply a Large Velocity Gradient modelling\cite{YoungScoville91} (LVG) to the SLED to gain insights into the physical properties of the gas, including its kinetic temperature and density \cite{Daddi15}.
We include in this modelling the CO and [CI] observations assuming ([CO]/[H2])/(dv/dr) = $10^{-5}$ and ([CI]/[H2])/(dv/dr) = $6 \times 10^{-6}$\cite{Weiss03}. 
To determine the best-fit model we use a customized $\chi^2$ minimization algorithm (MICHI2, \url{https://ascl.net/code/v/2533}, D. Liu et al. in prep).

{We note that a single component LVG model cannot reproduce well the observed CO/[CI] emission.
For the narrow emission the single component model yields a poor fit overall ($\chi^2_{red, single} = 7.22$ vs $\chi^2_{red, two} = 1.47$). In particular, this model significantly underestimates the CO(2-1) emission while over-predicting the CO(5-4) flux.
The broad emission might be formally fitted by a single component LVG model ($\chi^2_{\rm r, single} = 2.02 $ vs $\chi^2_{\rm r, two} = 1.99 $). 
However, the single component model underestimates the CO(5-4) flux and overestimates the [CI] emission by $\sim 1 \sigma$ respectively.
This is consistent with previous findings showing that at least two components are required to reproduce the SLEDs of high-redshift galaxies \cite{Daddi15}.
Therefore, we fit our data using a two-component LVG model.}

The CO emission from the galaxy (left panel in Figure \ref{Fig_LVGmodels}) shows enhanced CO(5-4) and CO(7-6) flux densities, indicative of large quantities of dense gas and highly excited interstellar medium. The low-excitation component of this gas has a density log(n$_{\rm H2}/{\rm cm}^{-3}) = 2.9 \pm 0.8$ whereas the high-excitation component has log(n$_{\rm H2}/{\rm cm}^{-3}) = 5.2 \pm 1.2$. 
These results are consistent with the starbursting nature of the galaxy and supports our choice of a ULIRG-like $\alpha_{\rm CO}$ conversion factor for computing the gas mass of the galaxy.
{The AGN is expected to contribute less than 1\% to the high-J CO luminosities given its X-ray luminosity and the galaxy SFR\cite{Spaans08}}.

The CO emission from the ejected material (right panel in Figure \ref{Fig_LVGmodels}) is instead characterized by an overall low CO excitation, at odds with what expected from a merging starburst. 
Merging starbursts in the local Universe are in fact characterized by a highly excited SLED overall\cite{Papadopoulos12} and the broad component is substantially less excited than the least excited of these objects. For example, the $R_{72} = L'_{\rm CO(7-6)}/L'_{\rm CO(2-1)}$ ratio measured in Arp193 (the least excited ULIRG in Reference \cite{Papadopoulos14}, see their Figure 17 and Table 1) is $\sim 4 \times$ higher than the upper limit measured in the broad component of ID2299 {. When accounting for measurements errors, the probability that these two ratios are consistent is $\sim 0.1 \%$.}
We find evidence of the presence of ISM-like low- and high-density gas having log(n$_{\rm H2}/{\rm cm}^{-3}) = 2.7 \pm 0.7$ and log(n$_{\rm H2}/{\rm cm}^{-3}) = 4.55 \pm 1.85$ respectively.
This is consistent with the molecular gas density properties of high-redshift disks\cite{Daddi15} containing both low- and moderately high-density gas and suggests that we are observing ISM-like material tidally stripped from the original merging galaxies. In the feedback-driven wind scenario we would instead expect highly excited, low density gas \cite{Weiss00}.

\subsection{ \underline{Disruptive events rate computation}}
\label{Sect_Rate}

We define the disruptive events rate as the ratio between the number of detected event in the comoving volume and the typical duration of the event. 
ID2299 is the only galaxy displaying evidence of an ongoing disruptive event in our ALMA survey and it is a starburst galaxy having a SFR significantly enhanced with respect to the average star-forming galaxy population at similar epoch. 
We observed 35 of such starburst galaxies in our ALMA survey {  where these are defined as galaxies having SFR$\geq 4 \times$ SFR$_{\rm MS}$}.
This results into a $3 \%$ disruptive events fraction among starbursts.
To obtain the total number of disruptive events in the comoving volume we include a correction factor accounting for the fact that we observe only the $7$ \% of starburst galaxies between $1.1 \leqslant z \leqslant 1.7$ (Ref. \cite{Jin18}). 
We further considered a typical disruptive event time-scale t$_{\rm disr} \sim {\rm R}_{\rm OF} / <{\rm v}_{\rm OF}> = 5 \pm 1 $ Myr. Here $<{\rm v}_{\rm OF}>$ is taken as the average between v$_{\rm max}$ and the velocity shift of the broad component v$_{\rm mol,b}$. Considering v$_{\rm max}$ or v$_{\rm mol,b}$ has no impact on the  disruptive events rate measurement
Finally, the cosmic volume sampled by our survey corresponds to $10^7$ Mpc$^3$. 
The error on the disruptive events rate is derived by adding in quadrature the Poissonian error associated to a single event observation and the uncertainties on the time-scale. 

{  We compare the disruptive event rate to the density of starburst galaxies producing a major ejection event. 
We expect in fact disruptive tidal ejections to correlate with compactification of the gas and thus with merger-driven starburst activity. This is because the nuclear starburst is fuelled by gas that loses its angular momentum while sinking into the center. For angular momentum conservation, a comparable amount of gas should thus be expelled from the system, and we can see it being ejected. 
For this computation we consider the space density of starburst galaxies in the cosmic volume sampled by our survey, computed as above as the number of galaxies with SFR$\geq 4 \times$ SFR$_{\rm MS}$ between $1.1 \leqslant z \leqslant 1.7$\cite{Jin18}.
However, not all starbursting mergers will result in a major ejection event, capable of quenching the galaxy. 
We thus correct for the 25\% fraction of starbursts that are expected to eject large tails of gas\cite{BournaudDuc06}.
We consider that the visibility window of the phenomenon is comparable to the dynamical time-scale of the merger t$_{\rm dyn, merger} = 100$ Myr \cite{Daddi10}.
We compute the errorbars by considering that the fraction of starbursts with large tails of gas varies between 20 and 30 \% whereas the visibility window varies between 50 and 200 Myr.}

\subsection{ \underline{Comparison with previous "disruptive events" observations}}

We dub ``disruptive events'' winds capable of expelling a substantial part of the galaxy ISM in the molecular phase. These events would have a direct and significant impact on the star formation of the host. 
Although molecular winds appear more massive than ionized ones \cite{HerreraCamus19}, previous analyses report that the molecular mass involved in feedback-driven outflows is typically one order of magnitude lower than the total molecular gas mass of the system \cite{Brusa18, Cicone12}

A disruptive event was previously reported for a $z \sim 0.7$ starburst galaxy with an optical size of R$_{\rm eff} \sim \ 94$ pc\cite{Geach14}. 
However, we argue here that this object has anomalous properties which make it not representative of the {\it massive} galaxies population.
The effective size of this galaxy is in fact about 9$\sigma$ below the typical size of star forming galaxies at $z \sim 0.75$ (R$_{\rm eff} \sim 5$ kpc  with $\sim 0.2 $ dex scatter, \cite{vanderWel14}). This galaxy seems even more compact than starburst galaxies at similar redshifts which hardly ever reach 1 kpc \cite{Calabro19}. 
This exceptionally small size would hint to a fairly low-mass galaxy, as also suggested by its M$_{\rm mol} \sim 3.1 \times 10^9 \ $ M$_{\odot}$ gas mass.  
However, the reported stellar mass of this object is large (M$_{\star} \sim 5.5 \times 10^{10} \ $ M$_{\odot}$) which would imply an anomalously low gas fraction. On the other hand, the dynamical mass derived from the galaxy size and CO velocity width reported would be in disagreement with the stellar mass reported. 

\subsection{ \underline{Comparison with local mergers with tidal tails}}

NGC3256 is a merging starbursting galaxy in the local Universe. This object has two nuclei close to coalescence, two extended tidal tails emitting in neutral hydrogen \cite{English03} and CO \cite{Aalto91} and two molecular winds arising from the two nuclei at small scales \cite{Sakamoto14}. 
Both phenomena produce broad wings around the nuclear emission (see Figure 1 in Reference \cite{Aalto91}, Figure 7 in \cite{Sakamoto06} and Figure 12 in \cite{Sakamoto14}). 
Deep and high resolution observations in the nuclear regions are essential to reveal the high velocity gas launched by the molecular wind (see e.g. Figure 1 in Ref. \cite{Sakamoto06}) and a comparison with the tidal tails optical morphology is crucial to disentangle the two phenomena\cite{Sakamoto14}. 
Observing this galaxy at $z \sim 1.4$ as in our observations, we would detect a spectrum with an overall broadening mostly produced by the tail, similar to what shown in Figure 1 of Reference \cite{Aalto91}. The wind would be indistinguishable from the tail emission since only a small part of the outflowing gas extends to very high velocities \cite{Sakamoto14}. 

NGC4194 is another example of a local (minor) merger having an extended tidal tail.
The narrow CO emission from the tail is shifted of $\sim 100$ km/s with respect to the nuclear emission and has a peak intensity that is 1/5 of that from the  galaxy nucleus\cite{Aalto01}.
Observing an analogous of this system at $z \sim 1.4$ with a $\sim 0.9''$ beam would show a single broad line skewed towards blue velocity. That is, this feature could not be formally distinguished from a wind and high spatial resolution observations would be required to work out the intrinsic nature of the phenomenon causing the line broadening.

{ NGC6240 is a nearby merger likely caught between the first encounter and the final coalescence stage\cite{Feruglio13}.
The bulk of the molecular gas resides in a bridge connecting the two pre-merger nuclei\cite{Tacconi99}.
While the dynamical status of this inter-nuclear gas is unclear\cite{Tacconi99,Feruglio13b,Cicone18}, the relatively low velocity and velocity dispersion (see e.g. Figure 3 in Reference \cite{Cicone18}) as well as the morphology connecting to a larger scale stream associated to a tidal tail \cite{Feruglio13} indicate a merger origin for this material. Such bridges of gas would be also expected at the merger stage of the source \cite{BarnesHernquist,ToomreToomre, Iono04}.
Another blue-shifted stream of gas is observed elongating in the east/west direction. 
The larger velocity shift and dispersion as well as some overlap with a highly shocked super-wind might suggest that this is a molecular outflow \cite{Feruglio13, Feruglio13b} although the large spatial extension would favour a merger-induced stream\cite{Tacconi99}. 
Finally, a small fraction of the molecular gas is likely expelled at high velocity by AGN/starburst activity from the nuclear regions\cite{Feruglio13b, Cicone18, Treistier20}.
When comparing NGC6240 with ID2299, there are key differences to consider. 
For example, unlike ID2299, NGC6240 hosts two AGN with quasar-like luminosities \cite{Cicone18}. NGC6240 also has a much lower SFR.
Furthermore, NGC6240 appears to be in an earlier stage of the interaction. 
Nonetheless, Figure 4 in Reference \cite{Cicone18} provides an example of how the NGC6240 spectrum might look like at high redshift.
This spectrum shows a narrow and a slightly red-shifted broad component, similar to that of our source. Hence, spatially integrated observations of NGC6240 seem to be dominated by the systemic and inter-nuclear emission, the latter likely arising from merger-induced gravitational disturbances. 
The contribution from the feedback-driven high velocity gas is marginal in this spectrum.}

\nolinenumbers

\begin{addendum}
 \item[Data Availability:] 
 The ALMA data analysed in this study are publicly available from the ALMA archive (\url{http://almascience.nrao.edu/aq/}, Program IDs: 2015.1.00260.S, 2016.1.00171.S, 2019.1.01702.S).
The DEIMOS spectrum of the source is also publicly available and can be retrieved through the COSMOS archive (\url{http://cosmos.astro.caltech.edu/}). 
 \item[Code Availability:] 
The ALMA data are processed using a series of GILDAS-based scripts available at \url{https://github.com/1054/Crab.Toolkit.PdBI}.
The GILDAS software is publicly available at \url{http://www.iram.fr/IRAMFR/GILDAS}. 
The CO/[CI] emission of the source have been modelled with the MICHI2 software which is publicly available at \url{https://ascl.net/code/v/2533}.

 \item[Competing Interests] The authors declare that they have no
competing financial interests.
\end{addendum}

\begin{figure}
\centering
\includegraphics[scale=0.7]{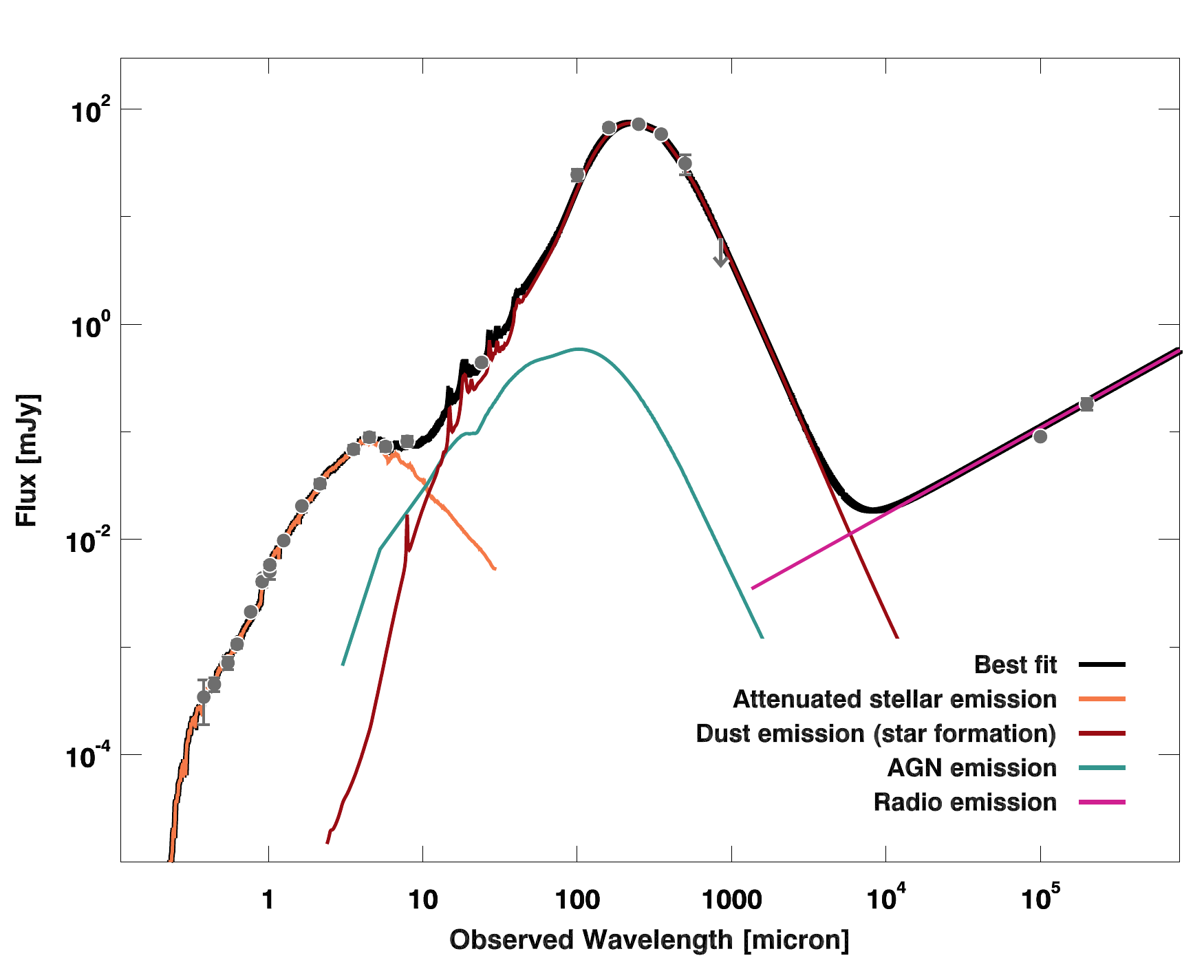}
\caption*{ {\bf Extended Data Figure 1: Spectral energy distribution of ID2299.} 
Dark grey dots represent the observed multi-wavelength photometry and 1$\sigma$ errors while dark grey arrows indicate $3 \sigma$ upper limits. 
The black solid line is the best-fit spectral energy distribution of the source (SED, see Methods for details). 
The coloured lines represent individual contributions to the best-fit SED from attenuated stellar emission (orange), dust emission from star formation (dark red), AGN (teal) and radio emission (magenta). }
\label{Fig2m_SED}
\end{figure}

\begin{figure}
\centering
\includegraphics[scale=0.7]{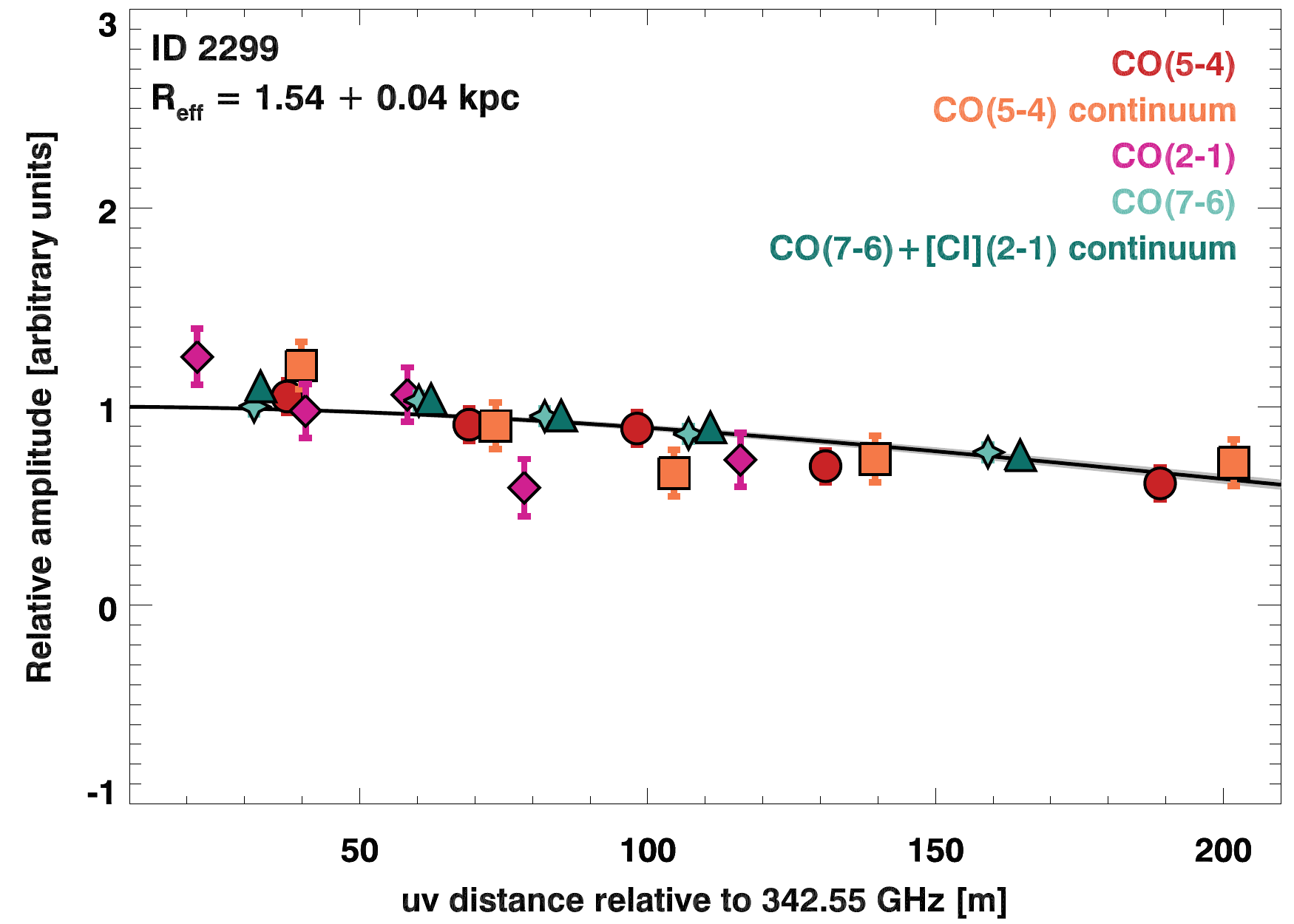}
\caption*{ {\bf Extended data Figure 2: Amplitude as a function of the \textit{uv} distance for ID2299. } 
Different symbols and colours show the amplitude and 1$\sigma$ error from the various transitions/continua (see legend).
The black line is the best-fit Gaussian profile. The grey shaded area highlights the 1$\sigma$ error associated to this model and it is comparable to the thickness of the black solid line.
The size of the best-fit Gaussian profile and 1$\sigma$ uncertainty are reported in the upper left corner of the plot. }
\label{Fig1m_amplitudes}
\end{figure}    

\newpage

%%%%%%%%%%%%%%%
%Bibliography

\end{document}